\documentclass[11pt,a4paper]{article}
\usepackage{amsmath,amssymb}
\usepackage[dvips]{graphicx}
\usepackage[nosort]{cite}

\allowdisplaybreaks

\begin{document}


\begin{center}

\hfill DESY 17-117\\

\vskip 1.25in

{\Large \bf 
Pole inflation in Jordan frame supergravity
}

\vskip 1.25in

{\bf Ken'ichi Saikawa$^{1}$, Masahide Yamaguchi$^{2}$, Yasuho Yamashita$^{3}$, and Daisuke Yoshida$^{4}$}
\vskip 0.25in

\begin{tabular}{lc}
&\!\! {$^{1}$ \em Deutsches Elektronen-Synchrotron DESY,}\\
&{\em Notkestrasse 85, 22607 Hamburg, Germany}\\[.4em]
&\!\! {$^{2}$ \em Department of Physics, Tokyo Institute of Technology,}\\
&{\em Ookayama, Meguro-ku, Tokyo 152-8551, Japan}\\[.4em]
&\!\! {$^{3}$ \em Yukawa Institute for Theoretical Physics, Kyoto University,}\\
&{\em Kyoto 606-8502, Japan}\\[.4em]
&\!\! {$^{4}$ \em Department of Physics, McGill University,}\\
&{\em Montr\'eal, QC, H3A 2T8, Canada}\\[.4em]

\end{tabular}

\vskip .5in

{\bf Abstract}
\end{center}

\begin{quote}
We investigate inflation models in Jordan frame supergravity, in which
an inflaton non-minimally couples to the scalar curvature. By imposing
the condition that an inflaton would have the canonical kinetic term in the Jordan
frame, we construct inflation models with asymptotically flat potential
through pole inflation technique 
and discuss their relation to the models based on Einstein frame supergravity.
We also show that the model proposed
by Ferrara {\it et al.} has special position and the relation between
the K\"ahler potential and the frame function is uniquely determined by requiring
that scalars take the canonical kinetic terms in the Jordan frame and that a
frame function consists only of a holomorphic term (and its
anti-holomorphic counterpart) for symmetry breaking terms. Our case
corresponds to relaxing the latter condition.
\end{quote}

\thispagestyle{empty}

\newpage

\tableofcontents

\section{Introduction}
\setcounter{equation}{0}

The observations of cosmic microwave background (CMB) radiation
anisotropies~\cite{Bennett:2012zja,Adam:2015rua} strongly support the
presence of inflation in the very early Universe and also constrain the
property of primordial perturbations. Though primordial tensor
perturbations have not yet been observed unfortunately, primordial
curvature perturbations are well constrained on CMB scales~\cite{Hinshaw:2012aka,Ade:2015lrj}. These constraints favor the
Starobinsky~\cite{Starobinsky:1980te} and the Higgs inflation models~\cite{CervantesCota:1995tz,Bezrukov:2007ep}, 
which suggest the $\alpha$-attractor models~\cite{Ferrara:2013rsa,Kallosh:2013yoa,Carrasco:2015pla} and the $\xi$-models
\cite{Pallis:2013yda,Giudice:2014toa,Pallis:2014dma,Kallosh:2014rha},
and these models are finally understood as pole inflation~\cite{Galante:2014ifa,Broy:2015qna} in a unified manner.
The pole inflation approach was further generalized in Ref.~\cite{Terada:2016nqg},
and applied to supergravity based models in Refs.~\cite{Nakayama:2016eqv,Kobayashi:2017qhk}.

The Higgs inflation model (and the Starobinsky model after introducing
the Lagrangian multiplier field) works well by introducing a non-minimal
coupling to the scalar curvature. Of course, by conformal transformation~\cite{Maeda:1988ab}, 
we can have an equivalent system, in which a scalar
field minimally couples to gravity. Thus, we can freely interchange the
Jordan frame with the Einstein frame as long as a conformal factor is positive definite
and there is no preferred frame in
principle~\cite{ArmendarizPicon:2002qb,Flanagan:2004bz,Catena:2006bd,Deruelle:2010ht,Chiba:2013mha,Kamenshchik:2014waa}. 
Then, one of strong motivations to introduce such a non-minimal coupling
(Jordan frame) is to easily realize (apparently) complicated potential
in the Einstein frame from simple potential in the Jordan frame~\cite{Futamase:1987ua}. 
Actually, in the Higgs inflation model, the
Higgs field has quartic power-law potential (in the large field
approximation) in the Jordan frame, which is converted to an
asymptotically flat potential in the Einstein frame.

This kind of analysis was done in the context of superconformal
supergravity. Though LHC has not yet found the supersymmetry
unfortunately, it is still expected to exist at the very high energy
scale where inflation happened because it can control quantum
corrections to an inflaton. Based on the superconformal supergravity
\cite{Kallosh:2000ve}, Ferrara {\it et al.} derived a $N = 1, d = 4$
supergravity action in the Jordan frame~\cite{Ferrara:2010yw}. They also
found~\cite{Ferrara:2010yw,Ferrara:2010in} that a special class of the
K\"ahler potential and the frame function gives rise to the canonical
kinetic terms of scalar fields in the Jordan frame. Based on this formalism,
they discussed various aspects of inflation
in the next-to-minimal supersymmetric Standard Model (NMSSM), which is
included in this class.

In this paper, we first show that the model proposed in
Refs.~\cite{Ferrara:2010yw,Ferrara:2010in} has special position and that the
relation between the K\"ahler potential and the frame function is uniquely
determined by requiring that scalars take the canonical kinetic terms in
the Jordan frame and the frame function consists only of a holomorphic
term (and its anti-holomorphic counterpart) except the conformal part.
Based on this observation, we relax the latter condition with keeping
the canonical kinetic term of scalar field in the Jordan frame and
construct inflation models with asymptotically flat potential in the
Einstein frame by use of pole inflation structure.

The organization of the paper is as follows. In the next section, after
briefly reviewing the Jordan frame supergravity obtained from adequate
gauge fixing on superconformal
supergravity~\cite{Ferrara:2010yw,Ferrara:2010in}, we explicitly show
that, in the model proposed by Ferrara {\it et al.}, the relation
between the K\"ahler potential and the frame function is uniquely
determined by imposing adequate conditions. In
Sec.~\ref{sec:inflation_models_in_J_frame_supergravity}, we review the
implications of this model and discuss its possible generalizations.  In
Sec.~\ref{sec:obtaining_inflation_potential}, after introducing the pole
inflation approach, we relax the condition proposed by
Refs.~\cite{Ferrara:2010yw,Ferrara:2010in}, and consider the
non-holomorphic frame function with keeping the canonical kinetic term
of scalar field in the Jordan frame. It is shown that even for such
relaxed conditions we can obtain asymptotically flat inflaton potential
through pole inflation technique.
The relation to the models based on Einstein frame supergravity is also discussed.
Final section is devoted to summary and discussions.

\section{Jordan frame supergravity}
\setcounter{equation}{0}

In this section, we briefly review the Jordan frame supergravity according to
Refs.~\cite{Ferrara:2010yw,Ferrara:2010in}.  After describing the
bosonic part of the supergravity action in the Jordan frame, we
explicitly show that the model proposed by Ferrara {\it et
al.}~\cite{Ferrara:2010yw} is special in that the relation between the
K\"ahler potential and the frame function is uniquely determined by
assuming that scalars take the canonical kinetic terms in the Jordan
frame and that the frame function consists only of a holomorphic term (and
its anti-holomorphic counterpart) except the conformal part.
Hereafter we take the unit of $M_{\rm Pl} =1$ unless otherwise stated,
where $M_{\rm Pl}$ is the reduced Planck mass.

\subsection{Supergravity action in the Jordan frame}

The action in the Jordan frame supergravity is specified by four
functions, the K\"{a}hler potential $\mathcal{K}(z,\bar{z})$, superpotential
$W(z)$, gauge kinetic function $f_{AB}(z)$, and frame function
$\Phi(z,\bar{z})$.  The bosonic part of the Lagrangian density in the Jordan
frame is given by~\cite{Ferrara:2010yw}
\begin{align}
\frac{\mathcal{L}}{\sqrt{-g_J}} &= -\frac{1}{6}\Phi \mathcal{R}_J + \left(\frac{1}{3}\Phi g_{\alpha\bar{\beta}}-\frac{\Phi_{\alpha}\Phi_{\bar{\beta}}}{\Phi}\right) g_J^{\mu\nu}\hat{\partial}_{\mu} z^{\alpha}\hat{\partial}_{\nu} \bar{z}^{\bar{\beta}} \nonumber\\
&\quad - \frac{1}{4\Phi}\left(\Phi_{\alpha}\hat{\partial}_{\mu} z^{\alpha} - \Phi_{\bar{\beta}}\hat{\partial}_{\mu} \bar{z}^{\bar{\beta}}\right)\left(\Phi_{\gamma}\hat{\partial}_{\nu} z^{\gamma} - \Phi_{\bar{\delta}}\hat{\partial}_{\nu} \bar{z}^{\bar{\delta}}\right)g_J^{\mu\nu} - V_J + \mathcal{L}_1,
\label{Lagrangian_J}
\end{align}
where
\begin{align}
\hat{\partial}_{\mu} z^{\alpha} = \partial_{\mu} z^{\alpha} - A_{\mu}^A k_A^{\alpha}
\end{align}
are gauge covariant derivatives of scalar fields $z^{\alpha}$ with $\alpha=1,\dots,n$
for the theory including $n$ supermultiplets,
$A_{\mu}^A$ are Yang-Mills gauge fields with $A$ representing the gauge indices,
$k_A^{\alpha}$ are Killing vectors of the K\"{a}hler manifold,
$\mathcal{R}_J$ is the scalar curvature obtained from the Jordan frame metric $g_{J\mu\nu}$,
\begin{equation}
g_{\alpha\bar{\beta}} = \frac{\partial}{\partial z^{\alpha}}\frac{\partial}{\partial \bar{z}^{\bar{\beta}}}\mathcal{K}(z,\bar{z})
\end{equation}
is the K\"{a}hler metric, and
\begin{align}
\Phi_{\alpha} \equiv \frac{\partial}{\partial z^{\alpha}}\Phi, \quad \Phi_{\bar{\beta}} \equiv \frac{\partial}{\partial \bar{z}^{\bar{\beta}}}\Phi.
\end{align}
Here, $\mathcal{L}_1$ contains the kinetic terms of gauge fields,
\begin{align}
\mathcal{L}_1 = -\frac{1}{4}(\mathrm{Re}f_{AB}) F^A_{\mu\nu} F^{\mu\nu B} + \frac{i}{4}(\mathrm{Im} f_{AB})F^A_{\mu\nu}\tilde{F}^{\mu\nu B},
\end{align}
which is conformal invariant and takes the same form both in the Jordan frame
and the Einstein frame. The scalar potential is given by
\begin{align}
V_J &= \frac{1}{9}\Phi^2 V_E, \\
V_E &= V_E^F + V_E^D, \\
V_E^F &= e^{\mathcal{K}}\left(-3 W\bar{W} + g^{\alpha\bar{\beta}}\nabla_{\alpha} W \nabla_{\bar{\beta}}\bar{W}\right), \\
V_E^D &= \frac{1}{2}(\mathrm{Re}f)^{-1\ AB}P_A P_B,
\end{align}
where
\begin{align}
\nabla_{\alpha} W &= \partial_{\alpha} W + (\partial_{\alpha}\mathcal{K})W, \\
P_A(z,\bar{z}) &= i(k_A^{\alpha}\partial_{\alpha}\mathcal{K}-r_A) = -i(k_A^{\bar{\alpha}}\partial_{\bar{\alpha}}\mathcal{K}-\bar{r}_A),
\end{align}
and the holomorphic function $r_A(z)$ becomes nonzero only if the
K\"{a}hler potential changes nontrivially under Yang-Mills gauge
transformations.

We can move into the Einstein frame by performing the following
reparameterization in Eq.~\eqref{Lagrangian_J}
\begin{align}
g_J^{\mu\nu} = \Omega^2 g_E^{\mu\nu}, \quad \Omega^2 = -\frac{1}{3}\Phi. \label{conformal_factor}
\end{align}
By using the following relations
\begin{align}
\sqrt{-g_J} &= \Omega^{-4}\sqrt{-g_E}, \\
\mathcal{R}_J &= \Omega^2\left(\mathcal{R}_E + 6\Box_E \ln\Omega -6g_E^{\mu\nu}\partial_{\mu}(\ln\Omega)\partial_{\nu}(\ln\Omega)\right), \\
\Box_E \ln\Omega &= \frac{1}{\sqrt{-g_E}}\partial_{\mu}\left(\sqrt{-g_E} g_E^{\mu\nu}\partial_{\nu}\ln\Omega\right),
\end{align}
where $\mathcal{R}_E$ is the scalar curvature obtained from the Einstein frame metric $g_{E\mu\nu}$,
we obtain the Lagrangian in the Einstein frame (up to the total derivative)
\begin{align}
\frac{\mathcal{L}}{\sqrt{-g_E}} = \frac{1}{2}\mathcal{R}_E - g_{\alpha\bar{\beta}} g_E^{\mu\nu}\hat{\partial}_{\mu} z^{\alpha}\hat{\partial}_{\nu} \bar{z}^{\bar{\beta}} - V_E + \mathcal{L}_1.
\label{Lagrangian_E}
\end{align}
Note that the Einstein frame action does not depend on the function $\Phi(z,\bar{z})$. The kinetic terms of scalar fields are solely determined by the K\"{a}hler potential $\mathcal{K}(z,\bar{z})$.

The complete $N=1$ supergravity action including fermions in an arbitrary Jordan
frame was obtained in Ref.~\cite{Ferrara:2010yw} by applying
superconformal approach to supergravity~\cite{Kallosh:2000ve}.  In this
approach, the Jordan frame action which is invariant under the
super-Poincar\'{e} algebra is derived by gauge fixing all extra
symmetries in the superconformal algebra. In this gauge-fixing
procedure, the basis of chiral supermultiplets $\{X^I\}$ in the
superconformal theory is changed into a basis $\{y,z^{\alpha}\}$, where
$y$ transforms under dilatations, but $z^{\alpha}$ do not transform
under them.  After applying the gauge condition, $y$ is no longer a
dynamical field and becomes a non-holomorphic function of other physical
scalar fields $z^{\alpha}$.  In particular, the gauge condition used in
Ref.~\cite{Ferrara:2010yw} reads
\begin{equation}
y = \bar{y} = \sqrt{-\frac{\Phi}{3}}\exp\frac{\mathcal{K}}{6}.
\end{equation}
This implies that $y$ and $\bar{y}$ are determined as a function of $z^{\alpha}$ and $\bar{z}^{\bar{\alpha}}$ once we specify $\mathcal{K}$ and $\Phi$.
It will be convenient to rewrite the above equation as
\begin{equation}
\mathcal{K}(z,\bar{z}) = -3\log\left(-\frac{1}{3}\frac{\Phi(z,\bar{z})}{Y(z,\bar{z})}\right), \label{K_Phi_relation_general}
\end{equation}
where we introduced a function $Y(z,\bar{z}) \equiv y(z,\bar{z})\bar{y}(z,\bar{z})$.

\subsection{FKLMP formulation}
\label{sec:FKLMP_formulation}

Ferrara, Kallosh, Linde, Marrani, and Van Proeyen
(FKLMP)~\cite{Ferrara:2010yw} proposed the following conditions to
obtain canonical kinetic terms for scalar fields in the Jordan frame:

\begin{enumerate}
\item The frame function and the K\"{a}hler potential have the following relation
\begin{align}
\mathcal{K}(z,\bar{z}) = -3\log\left(-\frac{1}{3}\Phi(z,\bar{z})\right).
\label{K_Phi_relation_FKLMP}
\end{align}
This condition corresponds to the fact that we arrange the K\"{a}hler potential such that $Y(z,\bar{z})$ becomes unity in Eq.~\eqref{K_Phi_relation_general}.
\item The frame function takes the following form
\begin{align}
\Phi(z,\bar{z}) = -3 + \delta_{\alpha\bar{\beta}} z^{\alpha}\bar{z}^{\bar{\beta}} + J(z) + \bar{J}(\bar{z}),
\label{condition_FKLMP_Phi}
\end{align}
where $J(z)$ is a holomorphic function, and the scalar field configuration satisfies the following condition
\begin{equation}
\mathcal{A}_{\mu} = 0.
\label{condition_FKLMP_A}
\end{equation}
Here, $\mathcal{A}_{\mu}$ is given by
\begin{equation}
\mathcal{A}_{\mu} = - \frac{i}{2\Phi}\left(\Phi_{\alpha}\hat{\partial}_{\mu} z^{\alpha} - \Phi_{\bar{\beta}}\hat{\partial}_{\mu} \bar{z}^{\bar{\beta}}\right),
\end{equation}
which corresponds to the auxiliary field in the superconformal formulation of supergravity.
\end{enumerate}

Indeed, if we substitute Eq.~\eqref{K_Phi_relation_FKLMP} into Eq.~\eqref{Lagrangian_J}, the Jordan frame Lagrangian becomes
\begin{align}
\mathcal{L}_{\rm FKLMP} = \sqrt{-g_J}\left[-\frac{1}{6}\Phi \mathcal{R}_J - \Phi_{\alpha\bar{\beta}}\hat{\partial}_{\mu} z^{\alpha}\hat{\partial}_{\nu} \bar{z}^{\bar{\beta}} g_J^{\mu\nu}
+\Phi \mathcal{A}_{\mu}^2 - V_J + \mathcal{L}_1\right].
\label{L_FKLMP}
\end{align}
Then it reduces to the form with canonical kinetic terms for scalar
fields when we apply conditions given by
Eqs.~\eqref{condition_FKLMP_Phi} and~\eqref{condition_FKLMP_A}.

It should be noticed, as mentioned in \cite{Ferrara:2010yw}, that this set
of the frame function and the K\"ahler potential (together with
$\mathcal{A}_{\mu} = 0$) is sufficient to give the canonical kinetic
terms of scalar fields in the Jordan frame.
These conditions, however, seem to be ad hoc, and there might be a room to extend,
since in general we can use Eq.~\eqref{K_Phi_relation_general} for the
K\"{a}hler potential rather than Eq.~\eqref{K_Phi_relation_FKLMP}.
Furthermore, it might be possible to use the following frame function,
\begin{equation}
\Phi(z,\bar{z}) = -3 + \delta_{\alpha\bar{\beta}}z^{\alpha}\bar{z}^{\bar{\beta}} + \mathcal{J}(z,\bar{z}),
\label{Phi_beyond_FKLMP}
\end{equation}
where $\mathcal{J}(z,\bar{z})$ is an arbitrary function of $z^{\alpha}$
and $\bar{z}^{\bar{\alpha}}$.
In Sec.~\ref{sec:inflation_models_in_J_frame_supergravity} we will argue that this $\mathcal{J}(z,\bar{z})$ term
can be identified as a deformation of the K\"ahler potential with a specific symmetric structure.
The FKLMP model corresponds to the case where the function
$\mathcal{J}(z,\bar{z})$ becomes a holomorphic form $J(z)+\bar{J}(\bar{z})$.

Here we explicitly show that the special relation between the K\"{a}hler
potential and the frame function [Eq.~\eqref{K_Phi_relation_FKLMP}]
follows from the requirement that the kinetic terms of the scalar fields
are canonically normalized and that the $\mathcal{J}(z,\bar{z})$ term in 
Eq.~\eqref{Phi_beyond_FKLMP} is given by a holomorphic function and its anti-holomorphic counterpart.

The kinetic term in the Jordan frame is given by\footnote{In addition to the
term shown in Eq.~\eqref{L_kin_J}, there exists a term of the form
$\Phi \mathcal{A}_{\mu}^2$, but here we do not consider this contribution since it vanishes for the
configuration $z^{\alpha}=\bar{z}^{\bar{\alpha}}$.  Such a configuration
might be guaranteed if the imaginary parts of the scalar fields are
heavy and stabilized at the origin during inflation.}
\begin{align}
\frac{\mathcal{L}_{{\rm kin},J}}{\sqrt{-g_J}} = \left(\frac{1}{3}\Phi g_{\alpha\bar{\beta}}-\frac{\Phi_{\alpha}\Phi_{\bar{\beta}}}{\Phi}\right) g_J^{\mu\nu}\hat{\partial}_{\mu} z^{\alpha}\hat{\partial}_{\nu}\bar{z}^{\bar{\beta}}.
\label{L_kin_J}
\end{align}
In order to have canonical kinetic terms, we require that
\begin{equation}
g_{\alpha\bar{\beta}} = \frac{3}{\Phi}\left(\frac{\Phi_{\alpha}\Phi_{\bar{\beta}}}{\Phi}-\delta_{\alpha\bar{\beta}}\right). \label{condition_canonical}
\end{equation}
From Eq.~\eqref{K_Phi_relation_general}, we also have
\begin{equation}
g_{\alpha\bar{\beta}} = -3 \left(\frac{\Phi_{\alpha\bar{\beta}}}{\Phi}-\frac{\Phi_{\alpha}\Phi_{\bar{\beta}}}{\Phi^2} 
- \frac{Y_{\alpha\bar{\beta}}}{Y}+\frac{Y_{\alpha}Y_{\bar{\beta}}}{Y^2}\right).
\label{Kahler _metric_Phi_Y}
\end{equation}
Combining these two equations, we obtain
\begin{equation}
\frac{\delta_{\alpha\bar{\beta}}}{\Phi} = \frac{\Phi_{\alpha\bar{\beta}}}{\Phi} 
- \frac{Y_{\alpha\bar{\beta}}}{Y}+\frac{Y_{\alpha}Y_{\bar{\beta}}}{Y^2}.
\label{condition_canonical_Y}
\end{equation}
If we choose the holomorphic term, $\mathcal{J}(z,\bar{z})=J(z)+\bar{J}(\bar{z})$,
Eq.~\eqref{condition_canonical_Y} reduces to
\begin{equation}
\left(\log Y\right)_{\alpha\bar{\beta}} = 0.
\label{condition_canonical_Y_holomorphic_J}
\end{equation}
Since the K\"{a}hler potential and hence $\log Y$ should be a real function,
the solution of Eq.~\eqref{condition_canonical_Y_holomorphic_J} is of the form
\begin{equation}
\log Y = h(z) + \bar{h}(\bar{z}) + \mathrm{const.},
\label{solution_canonical_Y_holomorphic_J}
\end{equation}
where $h(z)$ is some holomorphic function. The terms in the right-hand
side of Eq.~\eqref{solution_canonical_Y_holomorphic_J} can be eliminated
by using K\"{a}hler transformations.  Therefore, we must use the
relation given by Eq.~\eqref{K_Phi_relation_FKLMP} as long as we assume
that the kinetic terms are canonically normalized in the Jordan frame and
that the $\mathcal{J}(z,\bar{z})$ term is holomorphic.  Conversely, we expect
that the relation between $\mathcal{K}(z,\bar{z})$ and $\Phi(z,\bar{z})$
is modified as Eq.~\eqref{K_Phi_relation_general} once we consider some
non-holomorphic function for the $\mathcal{J}(z,\bar{z})$ term.

\section{Inflation models in Jordan frame supergravity}
\setcounter{equation}{0}
\label{sec:inflation_models_in_J_frame_supergravity}

The implications of the FKLMP formulation in the context of inflationary cosmology were discussed further in
Refs.~\cite{Ferrara:2010in,Kallosh:2013wya}. Before discussing such supergravity based models, let us first
consider the following (non-supersymmetric) toy model with two real scalar fields $\chi$ and
$\varphi$,
\begin{equation}
\mathcal{L} = \sqrt{-g}\left[\frac{1}{2}\partial_{\mu}\chi\partial_{\nu}\chi g^{\mu\nu} - \frac{1}{2}\partial_{\mu}\varphi\partial_{\nu}\varphi g^{\mu\nu} + \frac{\chi^2}{12}\mathcal{R} - \frac{\varphi^2}{12}\mathcal{R} - \frac{\lambda}{4}\varphi^4\right].
\label{L_compensator}
\end{equation}
$\chi$ has a wrong sign in the kinetic term, but it is not a problem
because $\chi$ is the compensator field rather than the physical field. 
The Lagrangian~\eqref{L_compensator} is
invariant under the following conformal transformations
\begin{align}
g_{\mu\nu} &\to e^{-2\sigma(x)}g_{\mu\nu}, \nonumber\\
\varphi &\to e^{\sigma(x)}\varphi, \nonumber\\
\chi &\to e^{\sigma(x)}\chi,
\label{conformal_transformations}
\end{align}
where $\sigma(x)$ is a position-dependent parameter.  If we take the
conformal gauge as $\chi = \sqrt{6}$, we obtain the Jordan frame Lagrangian
with the single scalar field $\varphi$,
\begin{align}
\mathcal{L} = \sqrt{-g}\left[\frac{1}{2}\mathcal{R} - \frac{1}{2}\partial_{\mu}\varphi\partial_{\nu}\varphi g^{\mu\nu} - \frac{\varphi^2}{12}\mathcal{R} - \frac{\lambda}{4}\varphi^4\right].
\label{L_compensator_fixed}
\end{align}

The above Lagrangian contains the non-minimal scalar-curvature coupling $\frac{\xi}{2}\varphi^2\mathcal{R}$ with $\xi = -\frac{1}{6}$.
On the other hand, in order to explain the observational results through the inflationary model with the quartic potential and non-minimal scalar-curvature coupling
$\frac{\xi}{2}\varphi^2\mathcal{R} - \frac{\lambda}{4}\varphi^4$, generically we need a value $\xi \ne -\frac{1}{6}$.
In fact, such a model leads to asymptotically flat potential in the Einstein frame, and it can be shown that a value of $\xi \gtrsim \mathcal{O}(10^{-3})$
is required to explain the observational results~\cite{Kallosh:2013wya}.
This fact implies that inflation cannot be explained in the above toy model with two real scalar fields and the quartic potential,
unless we introduce an explicit breaking of the conformal symmetry ($\xi \ne -\frac{1}{6}$).

As was argued in Ref.~\cite{Kallosh:2013wya}, the situation becomes different when we consider the superconformal supergravity models,
which contain complex scalars rather than real scalars.
In particular, it is possible to obtain a suitable non-minimal scalar-curvature coupling $\xi \ne -\frac{1}{6}$ only from
the spontaneous breaking of superconformal symmetry rather than breaking it explicitly.
Furthermore, the form of the non-minimal scalar-curvature coupling and the kinetic terms of scalar fields 
can be associated with the structure of the K\"ahler potential and the frame function,
and hence the construction of the inflaton potential becomes quite non-trivial according to the choice of them.
In the following subsections, we clarify implications of the FKLMP formulation by following the argument
in Ref.~\cite{Kallosh:2013wya}, and discuss its generalizations.

\subsection{Canonical superconformal supergravity models and their deformations}
\label{sec:CSS}

Let us consider a theory with one complex scalar field $\phi$ with the following choice of the K\"ahler potential
and the frame function,
\begin{align}
\mathcal{K} &= -3\log\left(-\frac{1}{3}\Phi\right), \nonumber\\
\Phi &= -3 + |\phi|^2 -3\Delta\left(\phi^2+\bar{\phi}^2\right),
\label{Phi_Delta_deformed_CSS}
\end{align}
where $\Delta$ is a real constant parameter.
The above choice corresponds to the FKLMP action [Eqs.~\eqref{K_Phi_relation_FKLMP}
and~\eqref{condition_FKLMP_Phi}] with $J(\phi) = -3\Delta \phi^2$.
This $\Delta$ term leads to the scalar-curvature coupling for the real and imaginary parts of $\phi$,
\begin{equation}
\frac{\mathcal{L}_J}{\sqrt{-g_J}} \supset \frac{\xi_{\rm r}}{2}\varphi_{\rm r}^2 \mathcal{R}_J + \frac{\xi_{\rm i}}{2}\varphi_{\rm i}^2 \mathcal{R}_J \quad \text{with} \quad 
\xi_{\rm r} = -\frac{1}{6} + \Delta \quad \text{and} \quad \xi_{\rm i} = -\frac{1}{6} - \Delta,
\end{equation}
where $\varphi_{\rm r} = \mathrm{Re}(\phi)/\sqrt{2}$ and $\varphi_{\rm i} = \mathrm{Im}(\phi)/\sqrt{2}$.

A special class of models obtained by setting $\Delta=0$ [or more generically $J(z) = 0$ in Eq.~\eqref{condition_FKLMP_Phi}]
is called the canonical superconformal supergravity (CSS) models~\cite{Ferrara:2010in}.
In the CSS models, both the real and imaginary parts are conformally coupled, {\it i.e.} $\xi_{\rm r} = \xi_{\rm i} = -\frac{1}{6}$.
Furthermore, it can be shown that the scalar potential in the Jordan frame becomes the same with that in the globally supersymmetric theory
if the superpotential consists only of cubic terms.
In the context of the superconformal construction of supergravity,
the CSS models emerge from a flat embedding K\"ahler manifold for chiral supermultiplets $\{X^I\}$
including the compensator field~\cite{Ferrara:2010in}.

Once we introduce $\Delta \ne 0$, both $\xi_{\rm r}$ and $\xi_{\rm i}$ deviate from the conformally invariant value $-\frac{1}{6}$.
For this reason, the terms proportional to $\Delta$, or more generically $\mathcal{J}(z,\bar{z})$ in Eq.~\eqref{Phi_beyond_FKLMP}
as well as $J(z)+\bar{J}(\bar{z})$, can be referred to as symmetry breaking terms~\cite{Ferrara:2010in}.
However, it should be noted that these terms do not explicitly break the underlying superconformal symmetry.
Actually, by arranging matter and compensator fields appropriately, it is possible to take $\Delta \ne 0$
with keeping the superconformal invariance of the theory, as explicitly shown in Ref.~\cite{Kallosh:2013wya}.\footnote{In other words, 
the scale transformations such as Eq.~\eqref{conformal_transformations}
do not correspond to dilatations in the superconformal theory.
The latter is defined in terms of the scale transformations for chiral supermultiplets $\{X^I\}$,
while only $y$ is changed under such transformations in the basis of $\{y,z^{\alpha}\}$ [see {\it e.g.} Ref.~\cite{Freedman:2012zz}].
Then, the relation between $\{X^I\}$ and $\{y,z^{\alpha}\}$ is arranged such that
\begin{equation}
X^I = yZ^I(z),
\end{equation}
where $Z^I(z)$ are holomorphic functions of $z^{\alpha}$.
The CSS models can be obtained by choosing $y=1$, $Z^0=\sqrt{3}$, and $Z^{\alpha}=z^{\alpha}$.}
Even though there is no explicit symmetry breaking term, the non-minimal scalar-curvature couplings
appear after imposing adequate gauge fixing conditions (or spontaneous breaking of the superconformal symmetry).
Furthermore, the embedding K\"ahler manifold deviates from flat geometry for $\Delta \ne 0$,
and its curvature does not vanish.
In this sense, $\Delta$ can be interpreted as a deformation parameter from the CSS models with 
a flat embedding K\"ahler manifold.

It is straightforward to extend the $\Delta$-deformed CSS models~\eqref{Phi_Delta_deformed_CSS}
to the case with a general holomorphic function $J(z)$ as in Eqs.~\eqref{K_Phi_relation_FKLMP} and~\eqref{condition_FKLMP_Phi}.
In particular, as long as we keep Eqs.~\eqref{K_Phi_relation_FKLMP} and~\eqref{condition_FKLMP_Phi},
the $F$-term scalar potential in the Jordan frame can be expressed as the following simple form~\cite{Buchmuller:2012ex}
\begin{align}
V_J^F &= \frac{1}{9}\Phi^2 e^{\mathcal{K}}\left(-3 W\bar{W} + g^{\alpha\bar{\beta}}\nabla_{\alpha} W \nabla_{\bar{\beta}}\bar{W}\right) \nonumber\\
& = V^{\rm global} + \Delta V_J,
\end{align} 
where 
\begin{equation}
V^{\rm global} = \delta^{\alpha\bar{\beta}}\partial_{\alpha}W\partial_{\bar{\beta}}\bar{W}
\end{equation}
is the $F$-term potential in the globally supersymmetric theory, and
\begin{equation}
\Delta V_J =  \frac{1}{(\Phi-\delta^{\gamma\bar{\delta}}\Phi_{\gamma}\Phi_{\bar{\delta}})}|\delta^{\alpha\bar{\beta}}\partial_{\alpha}W\Phi_{\bar{\beta}}-3W|^2.
\end{equation}
Then, there is a clear separation between the $F$-term potential in the global theory $V^{\rm global}$
and supergravity corrections $\Delta V_J$.
It can be shown that these supergravity corrections are at most comparable with $V^{\rm global}$
and that they become irrelevant under certain conditions~\cite{Das:2016kwz}.

\subsection{Non-holomorphic deformations}
\label{sec:generalize_CSS}

The FKLMP formulation can be regarded as a holomorphic deformation from the CSS models.
As shown in Sec.~\ref{sec:FKLMP_formulation}, this model has special point that
the relation between the K\"ahler potential and the frame function is uniquely
determined by requiring that scalars take the canonical kinetic terms in
the Jordan frame and that the frame function consists only of a holomorphic deformation term.
However, such a special relation does not hold in more generalized cases.
First, the additional terms in the frame function can be a non-holomorphic function $\mathcal{J}(z,\bar{z})$ as in Eq.~\eqref{Phi_beyond_FKLMP}.
Second, the relation between the K\"ahler potential and the frame function can be different from
Eq.~\eqref{K_Phi_relation_FKLMP}, and given by Eq.~\eqref{K_Phi_relation_general}
with some additional non-holomorphic function $Y(z,\bar{z})$.
In the remaining part of this paper, we consider such generalizations of the FKLMP formulation.

As a simple extension, we can consider the following non-holomorphic deformations of the CSS models,
\begin{align}
\mathcal{K} &= -3\log\left(-\frac{1}{3}\frac{\Phi}{Y}\right), \\
\Phi &= -3 + |\phi|^2 + \Delta_1 \mathcal{G}(\phi,\bar{\phi}), \\
Y&= 1 + \Delta_2 \mathcal{Y}(\phi,\bar{\phi}),
\end{align}
where $\mathcal{G}$ and $\mathcal{Y}$ are non-holomorphic functions,
and $\Delta_1$ and $\Delta_2$ are real constant parameters.\footnote{Non-holomorphic generalizations were also considered in Ref.~\cite{Nakayama:2010ga},
but the relation between the K\"ahler potential and the frame function was given by Eq.~\eqref{K_Phi_relation_FKLMP},
which corresponds to the case with $\Delta_2 = 0$.}
In this case, there might be no clear separation between the potential in the global theory and
supergravity corrections, and the kinetic terms of the scalar fields might deviate from the canonical form.

Even in such a complicated situation, we can still achieve some simplifications
if we introduce an additional field $S$ which takes $S=0$ during inflation, as shown in the next section and appendix~\ref{sec:non_holomorphic_Delta_CSS}.
For instance, if the superpotential takes the form $W = Sf(\phi)$,
the scalar potential in the Einstein frame can be written as [see Eq.~\eqref{V_E_non_holomorphic_Delta_CSS}]
\begin{equation}
V_E = \frac{(1+\Delta_2\mathcal{Y}(\phi,\bar{\phi}))^3}{(1-\frac{1}{3}|\phi|^2-\frac{\Delta_1}{3}\mathcal{G}(\phi,\bar{\phi}))^2}\left|f(\phi)\right|^2.
\end{equation}
If we further assume that $f$, $\mathcal{G}$, and $\mathcal{Y}$ are given by single power law functions,
at least it is possible to obtain an asymptotically constant scalar potential in the large field limit
by adjusting the powers of these functions.
However, it turns out that such a potential always exhibits runaway behavior in the large field limit, 
as explicitly shown in appendix~\ref{sec:non_holomorphic_Delta_CSS}, and
therefore it might not be suitable for inflation. 
A loophole in this argument is to consider the forms of the K\"ahler potential and/or the frame function 
that cannot be expressed in terms of single power law functions.
In the next section, we give an alternative framework to obtain the potential suitable for large field inflation
in the non-holomorphic extensions of the CSS models.

\section{Obtaining asymptotically flat inflaton potential}
\setcounter{equation}{0}
\label{sec:obtaining_inflation_potential}

A naive non-holomorphic extension of the CSS models in the previous section
revealed some difficulty in constructing the scalar potential suitable for inflation.
In this section, we follow a different approach, which takes advantage of the pole structure
of the kinetic term of the inflaton field in the Einstein frame~\cite{Galante:2014ifa,Broy:2015qna}.
Here we extend the FKLMP formulation by considering the non-holomorphic frame function,
but we still keep the canonical kinetic terms of scalar fields in the Jordan frame.
Combining the condition for the canonical kinetic terms in the Jordan frame with
the conditions for pole inflation, it becomes possible to constrain
the form of the K\"ahler potential and the frame function.
To this end, we first review the pole inflation approach proposed in Refs.~\cite{Galante:2014ifa,Broy:2015qna}
and reinterpret the FKLMP model in this framework in Sec.~\ref{sec:FKLMP_formulation_and_pole_inflation}.
After that, we apply this framework to the case with non-holomorphic generalizations
in Secs.~\ref{sec:beyond_FKLMP_formulation} and~\ref{sec:model_with_quadratic_frame_function}.
Finally, in Sec.~\ref{sec:relation_to_kinetic_formulation} we clarify the difference (or equivalence) between the models constructed in the Jordan frame
and those defined in the Einstein frame.

\subsection{FKLMP formulation and pole inflation}
\label{sec:FKLMP_formulation_and_pole_inflation}

In the NMSSM Higgs inflation considered in Refs.~\cite{Ferrara:2010yw,Ferrara:2010in}, 
the term of the form $J \propto H_u\cdot H_d$ was introduced for two Higgs
doublet fields $H_u$ and $H_d$ with keeping Eqs.~\eqref{K_Phi_relation_FKLMP} and~\eqref{condition_FKLMP_Phi}.
For large field values this term acts as a non-minimal scalar-curvature coupling, and the flat inflaton potential
can be realized. On the other hand, when the field value becomes
sufficiently small after inflation, the contribution from this term can
be neglected since it just leads to Planck-suppressed operators.

The flattening of the potential due to the non-minimal coupling was
reinterpreted in terms of the pole structure of the kinetic term of the
scalar field in the Einstein frame~\cite{Galante:2014ifa}. 
To highlight the essential points of this class of models, let us consider the
following Lagrangian for a real variable $\rho$ in the Einstein frame,
\begin{equation}\label{poleLag}
\mathcal{L} = \sqrt{-g}\left[\frac{1}{2}\mathcal{R} - \frac{1}{2}K_E(\rho)(\partial\rho)^2 - V_E(\rho)\right].
\end{equation}
Then, we adopt the following two assumptions:
\begin{enumerate}
\item $K_E(\rho)$ has a pole of order $p$ around $\rho\to 0$,
\begin{equation}
K_E(\rho) = \frac{a_p}{\rho^p} + \dots, \label{K_E_rho}
\end{equation}
where dots represent subleading terms, and $a_p$ is a constant parameter.
\item The scalar potential $V_E(\rho)$ is sufficiently smooth around the pole, namely,
\begin{equation}
V_E(\rho) = V_0(1-c\rho+\dots), \label{V_E_rho}
\end{equation}
with some constants $V_0$ and $c$.
\end{enumerate}
It is straightforward to compute the spectral index $n_s$, the
tensor-to-scalar ratio $r$, and the amplitude of the curvature
perturbations $\mathcal{P}_{\zeta}$, predicted by the above model for a number of e-foldings
$N$~\cite{Galante:2014ifa},
\begin{align}
n_s -1 &= -\frac{p}{(p-1)N}, \label{n_s_pole} \\ 
r &= \frac{8c^2}{a_p}\left(\frac{a_p}{c(p-1)N}\right)^{\frac{p}{p-1}},
\label{r_pole} \\ 
\mathcal{P}_{\zeta} &= \frac{V_E}{24\pi^2\epsilon},\label{amplitude_pole}
\end{align}
where $\epsilon \equiv (2K_E V_E^2)^{-1}(dV_E/d\rho)^2$ is the slow-roll parameter.
The spectral index is determined by the order of the pole $p$,
while the tensor-to-scalar ratio depends on other coefficients $a_p$
and $c$ as well as $p$.
In particular, for $p=2$ we obtain
\begin{equation}
n_s-1 = -\frac{2}{N},\quad r=\frac{8 a_2}{N^2}, \label{observables_p2}
\end{equation}
which is nicely fitted to the Planck results~\cite{Ade:2015lrj} for $a_2 \lesssim\mathcal{O}(1)$.

Let us reinterpret the FKLMP model with one complex scalar field $\phi$
and a quadratic form of $J(\phi)$ in terms of pole inflation.  From
Eqs.~\eqref{condition_FKLMP_Phi},~\eqref{condition_FKLMP_A}
and~\eqref{L_FKLMP}, we have
\begin{equation}
\mathcal{L} = \sqrt{-g_J}\left[-\frac{1}{6}\left(-3+|\phi|^2+J(\phi)+\bar{J}(\bar{\phi})\right)\mathcal{R}_J-\partial_{\mu}\phi\partial_{\nu}\bar{\phi}g^{\mu\nu}_J - V_J(\phi)\right],
\end{equation}
where we ignored the contribution from vector fields $\mathcal{L}_1$.
Introducing the deformation term $J(\phi)= -3\Delta\phi^2$ with $\Delta = \frac{1}{6}+\xi$ and assuming that the inflation occurs in the
trajectory with $\phi = \bar{\phi}=\varphi/\sqrt{2}$, we rephrase the
above theory in terms of the following single field model with a
non-minimal coupling to gravity,
\begin{equation}
\mathcal{L} = \sqrt{-g_J}\left[\frac{1}{2}\left(1+\xi\varphi^2\right)\mathcal{R}_J-\frac{1}{2}\partial_{\mu}\varphi\partial_{\nu}\varphi g^{\mu\nu}_J - V_J(\varphi/\sqrt{2})\right].
\label{L_FKLMP_J}
\end{equation}
After the reparameterization $g_J^{\mu\nu} =
(1+\xi\varphi^2)g_E^{\mu\nu}$, the Einstein frame Lagrangian reads
\begin{align}
\mathcal{L} = \sqrt{-g_E}\left[\frac{1}{2}\mathcal{R}_E - \frac{1}{2}K_E(\varphi)(\partial_{\mu}\varphi)^2 - \frac{V_J(\varphi/\sqrt{2})}{(1+\xi\varphi^2)^2}\right],
\end{align}
where
\begin{equation}
K_E(\varphi) = \frac{(1+\xi\varphi^2)+6\xi^2\varphi^2}{(1+\xi\varphi^2)^2}.
\end{equation}
If we define a new variable as
\begin{equation}
\rho \equiv (1+\xi\varphi^2)^{-1},
 \label{condition_FKLMP_rho}
\end{equation}
the kinetic term becomes
\begin{align}
K_E(\varphi)(\partial\varphi)^2 &= \left[\rho^2 + \left(\frac{1}{4\xi}+\frac{3}{2}\right)\frac{(\rho')^2}{\rho^2}\right]\frac{(\partial\rho)^2}{(\rho')^2},
\end{align}
where the prime represents the derivative with respect to $\varphi$.
Hence we see that the kinetic term has a second order pole at $\rho\to 0$ with a residue
\begin{equation}
a_2 = \frac{1}{4\xi}+\frac{3}{2}.
\end{equation}

The attractor-like predictions for inflationary observables $n_s$ and
$r$ can be obtained if the scalar potential has an appropriate form
around the pole.  Here it is sufficient to consider the quartic scalar
potential in the Jordan frame
\begin{equation}
V_J = \frac{\lambda}{4}\varphi^4, \label{V_FKLMP_J}
\end{equation}
where $\lambda$ is a real constant parameter.
In this case, the scalar potential in the Einstein frame becomes sufficiently smooth around the pole:
\begin{align}
V_E &= \frac{\lambda}{4}\frac{\varphi^4}{(1+\xi\varphi^2)^2} \xrightarrow{\rho\to 0} \frac{\lambda}{4\xi}(1-2\rho+\dots).
\end{align}
According to the general formulae~\eqref{n_s_pole} and~\eqref{r_pole}, the observational predictions of this model
are given by
\begin{equation}
n_s -1 = -\frac{2}{N},\quad r = \frac{8}{N^2}\left(\frac{1}{4\xi}+\frac{3}{2}\right).
\end{equation}
For $\xi\gg 1$, the predictions for $n_s$ and $r$ rarely depend on model parameters and converge
to the values $n_s = 1-2/N$ and $r = 12/N^2$.

We emphasize that the pole inflation approach guarantees just the
asymptotic flatness of the inflaton potential but not necessary a
graceful exit.  This is because the behavior in the small field regime
must be specified in subleading terms in Eqs.~\eqref{K_E_rho}
and~\eqref{V_E_rho}.  Although there are universal predictions for
inflationary observables due to the leading pole structure, the form of
the potential in the small field regime is model-dependent, and
inflation cannot end if the potential does not take an appropriate form
in the small field limit.  For instance, the potential may exhibit
runaway behavior and/or have a bump in the small field regime, which
might not be suitable for the reheating after inflation and/or
terminating the inflationary stage.\footnote{Of course, even if a
potential exhibits runaway behavior, efficient gravitational production
of particles~\cite{Ford:1986sy,Peebles:1998qn} or particle production via adiabaticity violation~\cite{Felder:1999pv}
might reheat the Universe. Also, even if a potential has a
bump in one direction, another direction might show waterfall behavior
and help an inflation terminate like hybrid inflation.} 
Therefore, it is necessary to check the full form of the potential in order to
guarantee the graceful exit as well as the asymptotic flatness of the
potential.

\subsection{Beyond FKLMP formulation}
\label{sec:beyond_FKLMP_formulation}

In the previous subsection, we have seen that a pole structure leading to
universal predictions for inflationary observables appears because of
the existence of the non-minimal scalar-curvature coupling which becomes
dominant in the large field regime.

The FKLMP formulation explicitly realizes this mechanism by introducing
a holomorphic term $J(z)$ in the frame function.
However, as discussed in Sec.~\ref{sec:generalize_CSS},
there might be a room to extend the choice of the K\"ahler potential and the frame function
in Eqs.~\eqref{K_Phi_relation_FKLMP} and~\eqref{condition_FKLMP_Phi} by introducing some non-holomorphic functions.
In the following, we aim to extend the FKLMP
formulation to construct pole inflation models by specifying the non-holomorphic function
$\mathcal{J}(z,\bar{z})$ in Eq.~\eqref{Phi_beyond_FKLMP} as well as the K\"{a}hler potential and
the superpotential.

For simplicity, here we focus on the models with two superfields $\phi$ and $S$,
and consider the following forms for the K\"{a}hler potential, frame function, and superpotential:
\begin{align}
\mathcal{K} &= \mathcal{K}\left(\phi+\bar{\phi},S\bar{S},S^2,\bar{S}^2\right), \label{K_function}\\
\Phi &= \Phi\left(\phi+\bar{\phi},S\bar{S},S^2,\bar{S}^2\right), \label{Phi_function}\\
W &= Sf(\phi). \label{W_function}
\end{align}
The second superfield $S$ is called the stabilizer field, which is introduced in order to avoid 
the negative contribution to the scalar potential~\cite{Kawasaki:2000yn}.
We also assume that $\mathcal{K}$ and $\Phi$ depend on the combinations
$\phi+\bar{\phi}$, $S\bar{S}$, $S^2$ and $\bar{S}^2$, which can be guaranteed by imposing 
shift symmetry $\phi\to\phi+i\mathcal{C}$ with $\mathcal{C}$ being a real constant parameter
and $\mathbb{Z}_2$ symmetry $S\to -S$~\cite{Kallosh:2010xz,Roest:2013aoa}.
The direction with $\mathrm{Re}(\phi) = (\phi+\bar{\phi})/2$ and $\mathrm{Im}(\phi)=S=0$
will play the role of the inflaton, which may have a non-canonical kinetic term leading to pole inflation in the Einstein frame.
However, we can still keep canonical kinetic terms in the Jordan frame by adjusting the frame function $\Phi$ and
the function $Y$, namely, the relation between $\Phi$ and $\mathcal{K}$.

We note that many simplifications are achieved when we use functions given by Eqs.~\eqref{K_function}-\eqref{W_function}.
In particular, if the field $S$ is stabilized at the origin during inflation, the scalar potential in the Einstein frame becomes
\begin{equation}
V_E = e^{\mathcal{K}}g^{S\bar{S}} |f|^2. \label{V_E_stabilized}
\end{equation}
Furthermore, if we require that kinetic terms of $\phi$ and $S$ are canonically normalized
in the Jordan frame [see Eq.~\eqref{condition_canonical}], we have
\begin{equation}
g^{S\bar{S}} = -\frac{\Phi}{3} \label{gSS_stabilized}
\end{equation}
at $S=0$. Therefore, in this case the scalar potential is simplified as
\begin{equation}
V_E = -\frac{\Phi}{3} e^{\mathcal{K}} |f|^2. \label{potential_simplified}
\end{equation}

In the following, we search for functions $\mathcal{K}$, $\Phi$, and $f$ which
realize pole inflation in the large field regime.
To construct the models, we adopt three guiding principles:

\begin{enumerate}

\item The scalar fields $\phi$ and $S$ should have canonical kinetic terms in the Jordan frame.
As shown in Eq.~\eqref{condition_canonical}, we require the following conditions\footnote{Note that
$g_{\phi \bar{S}} = g_{S\bar{\phi}}=0$ and $\Phi_S = \Phi_{\bar{S}}=0$ are automatically satisfied
at $S=0$, if $\mathcal{K}$ and $\Phi$ are given by Eqs.~\eqref{K_function} and~\eqref{Phi_function}, respectively.}
\begin{align}
g_{\phi\bar{\phi}} &= \frac{3}{\Phi}\left(\frac{\Phi_{\phi}\Phi_{\bar{\phi}}}{\Phi} - 1\right), \label{condition_canonical_phi}\\
g_{S\bar{S}} &= -\frac{3}{\Phi}, \label{condition_canonical_S}
\end{align}
at $S=0$. If the inflaton direction is identified as $\phi=\bar{\phi}=\varphi/\sqrt{2}$
with $\varphi$ being a real field, we have
\begin{equation}
\Phi_{\phi} = \Phi_{\bar{\phi}} = \frac{1}{\sqrt{2}}\Phi',
\end{equation}
and Eq.~\eqref{condition_canonical_phi} can be rewritten as
\begin{equation}
g_{\phi\bar{\phi}} = \frac{3}{\Phi}\left(\frac{(\Phi')^2}{2\Phi} - 1\right), \label{condition_canonical_varphi}
\end{equation}
where the prime represents the derivative with respect to $\varphi$.

\item The kinetic term of the inflaton field in the Einstein frame should have a specific structure leading to
pole inflation with the order of the pole $p$:
\begin{equation}
-\frac{1}{2}g_{\phi\bar{\phi}}(\partial\varphi)^2 \xrightarrow{\rho\to 0} -\frac{1}{2}\frac{a_p}{\rho^p}(\partial\rho)^2, 
\end{equation}
where the variable $\rho$ should be chosen such that it approaches $\rho\to 0$ for the large field regime,
and $a_p$ is a real constant [cf. Eq.~\eqref{K_E_rho}].
The above equation can be rewritten as
\begin{equation}
g_{\phi\bar{\phi}} \xrightarrow{\rho\to 0} \frac{a_p}{\rho^p}(\rho')^2. \label{condition_pole}
\end{equation}

\item The scalar potential in the Einstein frame should be sufficiently smooth at $\rho\to 0$:
\begin{align}
V_E &= -\frac{\Phi}{3} e^{\mathcal{K}} |f|^2 \xrightarrow{\rho\to 0} V_0(1-c\rho+\dots), \label{condition_potential}
\end{align}
where $V_0$ 
and $c$ are some constants.

\end{enumerate}

At this stage we have four unknown functions $\mathcal{K}$, $\Phi$, $f$, and $\rho$.
Once we specify the relation between 
two of them,
we expect that 
remaining three functions
are determined
by adopting three conditions described above.
Note that, however, we obtain the following nonlinear differential equation from Eqs.~\eqref{condition_canonical_varphi} and~\eqref{condition_pole},
\begin{equation}
\left(\frac{\partial \Phi}{\partial \rho}\right)^2 - \frac{2\Phi}{(\rho')^2} = \frac{2a_p\Phi^2}{3\rho^p}, \label{equation_Phi_rho}
\end{equation}
which cannot always be solved analytically.
In the following subsection, we give a concrete example, in which the above equation is solved analytically with 
an appropriate Ansatz.

\subsection{Model with a quadratic frame function}
\label{sec:model_with_quadratic_frame_function}
First of all, we need to specify the relation between two of unknown functions.
Here, we assume that the frame function approaches the following form,\footnote{It is also possible to consider the form
\begin{equation}
\Phi \xrightarrow{\rho\to 0} -A\rho^{-q} \nonumber
\end{equation}
instead of Eq.~\eqref{Phi_rho_relation},
where $q$ is a real positive constant. This choice corresponds to Eq.~\eqref{rho_Omega_relation}.
In this case, results for inflationary observables can be different up to the choice of the function $f$.
We will further discuss this point in Sec.~\ref{sec:relation_to_kinetic_formulation}.}
\begin{equation}
\Phi \xrightarrow{\rho\to 0} -A\rho^{-2},
\label{Phi_rho_relation}
\end{equation}
where $A$ 
is a real positive constant.
Next, in order to solve Eq.~\eqref{equation_Phi_rho}, we adopt the following Ansatz,
\begin{equation}
\frac{(\Phi')^2}{2\Phi} \xrightarrow{\rho\to 0} B,
\label{Phi_ansatz}
\end{equation}
where $B$ is some real constant.

Applying Eqs.~\eqref{Phi_rho_relation} and~\eqref{Phi_ansatz} to Eq.~\eqref{equation_Phi_rho}, we have
\begin{equation}
(\rho')^2 = \frac{3(1-B)}{a_p}\frac{\rho^{p+2}}{A}.
\end{equation}
By integrating the above equation, we obtain the relation between $\rho$ and $\varphi$,
\begin{equation}
\frac{\varphi}{\sqrt{2}} + C = \pm \sqrt{\frac{A a_p}{6(1-B)}}\frac{2}{p}\rho^{-\frac{p}{2}},
\end{equation}
where $C$ is an integration constant.
Substituting it into Eq.~\eqref{Phi_rho_relation}, we find
\begin{equation}
\Phi = \left\{
\begin{array}{ll}
\displaystyle -\frac{6(1-B)}{a_2}\left(\frac{\varphi}{\sqrt{2}}+C\right)^2 & \text{for}\quad p=2, \\
\displaystyle -A\left(\frac{3(1-B)p^2}{2Aa_p}\left(\frac{\varphi}{\sqrt{2}}+C\right)^2\right)^{\frac{2}{p}} & \text{for}\quad p\ne2.
\end{array}
\right. \label{Phi_limit}
\end{equation}
Hereafter, let us focus on the case with $p=2$. We can check that the above solution is consistent with
the Ansatz~\eqref{Phi_ansatz}, and the value of $B$ can be fixed as
\begin{equation}
B = \frac{(\Phi')^2}{2\Phi} = \frac{\frac{6}{a_2}}{\frac{6}{a_2}-1}.
\end{equation}
Then, the frame function becomes
\begin{equation}
\Phi = - \frac{\frac{6}{a_2}}{1- \frac{6}{a_2}}C^2 \left(1+\frac{\varphi}{\sqrt{2}C}\right)^2. \label{Phi_limit2}
\end{equation}

Equation~\eqref{Phi_limit2} characterizes the non-minimal scalar-curvature coupling in the large field limit ($\rho\to 0$).
However, the full structure of the frame function $\Phi$ must contain the term specifying the graviton kinetic term in the Jordan frame in addition to
the scalar-curvature coupling term obtained above.
Namely, we require that the frame function $\Phi$ should approach $\Phi\to -3$ in the small field limit.
This requirement can be satisfied by fixing the value of $C$,
\begin{equation}
C^2 = 3\left(\frac{a_2}{6}-1\right). \label{C_a2_relation}
\end{equation}
After applying this condition, we obtain
\begin{equation}
\Phi = -3\left(1+\sqrt{\frac{\tilde{\xi}}{3}}\varphi\right)^2, \label{Phi_tilde_xi}
\end{equation}
where we introduced a new parameter, 
\begin{equation}
\tilde{\xi} \equiv \frac{3}{2C^2}.
\end{equation}
Note that the parameter $\tilde{\xi}$ is related to $a_2$ via Eq.~\eqref{C_a2_relation},
\begin{equation}
a_2 = 6\left(1+\frac{1}{2\tilde{\xi}}\right).
\label{a2_tildexi_relation}
\end{equation}

Substituting the frame function~\eqref{Phi_tilde_xi} into the condition for the canonical kinetic term in the Jordan frame
\begin{equation}
(\log Y)'' = \frac{\Phi''-2}{\Phi}, \label{condition_canonical_kinetic_Y}
\end{equation}
we obtain
\begin{equation}\label{Y_quad_model}
\log Y = -2\left(1+\frac{1}{\tilde{\xi}}\right)\log\left(1+\sqrt{\frac{\tilde{\xi}}{3}}\varphi\right) + k_1\varphi + k_2.
\end{equation}
Hereafter, we drop the terms proportional to integration constants $k_1$ and $k_2$, since they can be shifted away by using K\"{a}hler transformations.
Then, the K\"{a}hler potential becomes
\begin{equation}
\mathcal{K} = -12\left(1+\frac{1}{2\tilde{\xi}}\right)\log\left(1+\sqrt{\frac{\tilde{\xi}}{3}}\varphi\right).
\end{equation}
Finally, the scalar potential is given by
\begin{align}
V_E &= -\frac{\Phi}{3}e^{\cal K}|f|^2 
= \left(1+\sqrt{\frac{\tilde{\xi}}{3}}\varphi\right)^{-10-\frac{6}{\tilde{\xi}}}|f|^2.
\end{align}

If we introduce the following superpotential
\begin{equation}
W = \lambda S\phi^m, 
\end{equation}
which corresponds to  $f = \lambda \phi^{m}$ with $m=5+3/\tilde{\xi}$ and $\lambda$ being a complex parameter,
the scalar potential becomes sufficiently smooth in the large field limit,
\begin{align}
V_E &= V_0\left(1+\frac{\sqrt{3}}{\sqrt{\tilde{\xi}}\varphi}\right)^{-2m}\nonumber\\
&\to V_0\left[1-\frac{2\sqrt{3}m}{\sqrt{\tilde{\xi}}\varphi}+\dots\right]\nonumber\\
&\to V_0\left[1-2m\sqrt{\frac{3(1+2\tilde{\xi})}{A}}\rho+\dots\right],
\label{V_E_quad_model}
\end{align}
where dots represents terms of higher order in $\sqrt{3}/\sqrt{\tilde{\xi}}\varphi$ or $\rho$, and
\begin{equation}\label{V_0}
V_0= |\lambda|^2 \left(\frac{3}{2\tilde{\xi}}\right)^m.
\end{equation}

It is possible to rewrite the K\"{a}hler potential, frame function, and superpotential for this model in terms of the complex scalar field $\phi$,
\begin{align}
\mathcal{K} &= -12\left(1+\frac{1}{2\tilde{\xi}}\right)\log\left(1+\sqrt{\frac{\tilde{\xi}}{6}}\left(\phi+\bar{\phi}\right)\right), \label{Kahler_quad_model} \\
\Phi &= -3 + |\phi|^2 + \mathcal{J}(\phi,\bar{\phi}), \\
W &= \lambda S\phi^m,
\end{align}
with\footnote{Alternatively, one can choose
\begin{equation}
\mathcal{J}(\phi,\bar{\phi}) = -\sqrt{6\tilde{\xi}}\left(\phi+\bar{\phi}\right) - (1+2\tilde{\xi})|\phi|^2.
\end{equation}
Note that $\mathcal{J}(\phi,\bar{\phi})$ is non-holomorphic, since it contains the term proportional to $|\phi|^2$.}
\begin{align}
m &= 5 + \frac{3}{\tilde{\xi}}, \\
\mathcal{J}(\phi,\bar{\phi}) &= -\sqrt{6\tilde{\xi}}\left(\phi+\bar{\phi}\right) - |\phi|^2 - \tilde{\xi} \left(\phi^2+\bar{\phi}^2\right). \label{non_horomorphic_J_quad_model}
\end{align}
From Eqs.~\eqref{observables_p2} and~\eqref{a2_tildexi_relation}, we estimate predictions for inflationary observables as
\begin{equation}
n_s-1 = -\frac{2}{N},\quad r = \frac{48}{N^2}\left(1+\frac{1}{2\tilde{\xi}}\right),\quad \mathcal{P}_{\zeta} = \frac{|\lambda|^2\tilde{\xi} N^2}{36\pi^2(1+2\tilde{\xi})}\left(\frac{3}{2\tilde{\xi}}\right)^m.
\label{predictions_observation}
\end{equation}
We present the plot of $n_s$ and $r$ for $N=50,\,60$ in Fig.~1. We see that $n_s$ and $r$
approach the values shown in the above equations for $\tilde{\xi}\gg 1$.
For the sake of comparison, we also show the prediction of the FKLMP model in the same plot.

\begin{figure}
  \begin{center}
   \includegraphics[width=150mm]{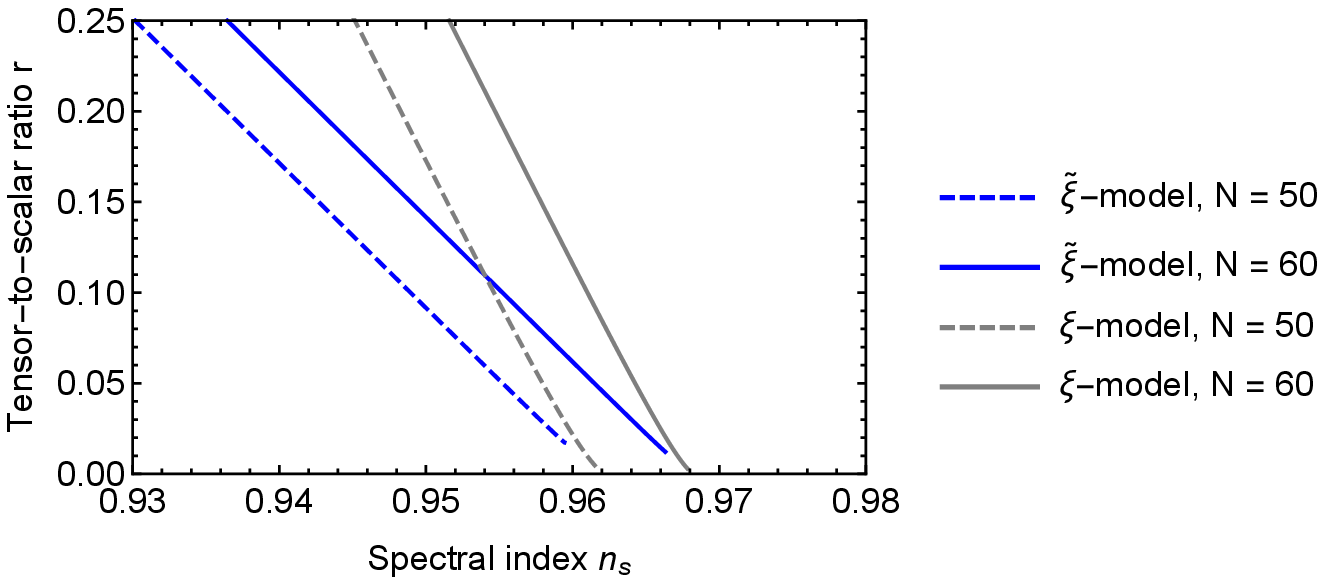}\\
   \vspace{10mm}
   \includegraphics[width=150mm]{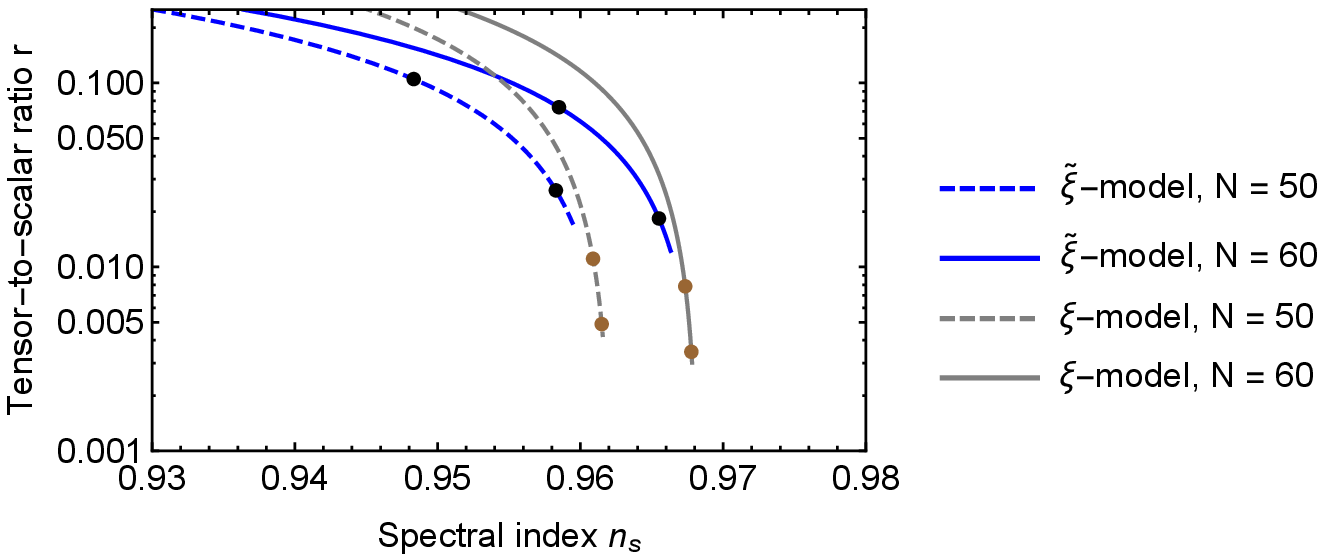}
  \end{center}
  \caption{
  In the both figures, blue and gray lines show the cosmological observables $n_s$ and $r$ for 
  the inflation model defined by Eqs.~\eqref{Kahler_quad_model}-\eqref{non_horomorphic_J_quad_model}
  (dubbed as $\tilde{\xi}$-model) and that defined by Eqs.~\eqref{L_FKLMP_J} and~\eqref{V_FKLMP_J} (dubbed as $\xi$-model), respectively.
  Solid (dashed) lines correspond to the predictions for $N=60\, (50)$.
  The black (brown) dots on each line in the bottom panel
  correspond to $\tilde{\xi}$ ($\xi$) $= 0.1$ and $1$ from top to bottom.}
  \label{fig1}
\end{figure}

In appendix~\ref{sec:stabilization}, we show that Eqs.~\eqref{Kahler_quad_model}-\eqref{non_horomorphic_J_quad_model}
are compatible with the assumption that $S$ and the imaginary part of $\phi$
are stabilized at the origin during inflation, if we introduce an appropriate higher order 
term of $S$ in the frame function.

We note that the potential~\eqref{V_E_quad_model} does not exhibit runaway behavior in contrast to 
a naive non-holomorphic extension of the CSS models discussed in Sec.~\ref{sec:generalize_CSS}.
The inflaton field rolls towards $\varphi\to 0$, and we expect that inflation ends at a point when the slow-roll condition is violated.

There are some remarks on the difference in comparison with the FKLMP model.
First, although $\Phi$ is quadratic in $\varphi$, the K\"{a}hler
potential has a non-trivial structure due to the existence of
non-holomorphic terms, and the additional factor
$V_0\propto \tilde{\xi}^{-m}$ is multiplied in the scalar potential. Because
of this factor, the amplitude of the curvature perturbation is
suppressed as $\mathcal{P}_{\zeta}\propto \tilde{\xi}^{-5+3/\tilde{\xi}}$, 
which may be fitted
to the observational result $\mathcal{P}_{\zeta}\simeq
\mathcal{O}(10^{-10})$ for $\tilde{\xi}\simeq \mathcal{O}(10^2)$
and $\lambda\simeq \mathcal{O}(1)$.  On the other hand, the prediction
of the FKLMP model reads $\mathcal{P}_{\zeta}\propto \xi^{-2}$, which
requires the value of $\xi$ as large as $\mathcal{O}(10^5)$ provided
that $\lambda\simeq \mathcal{O}(1)$. 
If we require that $m$ should be positive integer, $\tilde{\xi}$ is uniquely
determined to be either $1$ or $3$, which leads to $m = 8$ or $6$,
respectively. In this case, $\tilde{\xi}$ is of the order of unity though
$\lambda$ must be $\mathcal{O}(10^{-5})$.
Second, the power law superpotential $W \propto S\phi^m$ is required in
order to compensate the non-trivial $\varphi$ dependence of $V_E$
arising from the K\"{a}hler potential. In general, the power $m$ is not
integer number.
Finally, both in the FKLMP model and in this model, the prediction for the tensor-to-scalar ratio
has a lower limit, but there is a difference in its value.
In the FKLMP model the lower limit reads
$r \gtrsim 12/N^2$ for $\xi\gg1$, while in the above
model the lower limit reads $r \gtrsim 48/N^2$ for $\tilde{\xi}\gg 1$.

\subsection{Relation to kinetic formulation in the Einstein frame}
\label{sec:relation_to_kinetic_formulation}
In this subsection, we clarify the origin of differences or similarities of the models considered in this paper,
and discuss their relation to the models constructed in the Einstein frame.
First, we review the difference between the FKLMP model and the models constructed in the previous subsection
by the use of the variable $\rho$ introduced in Sec.~\ref{sec:FKLMP_formulation_and_pole_inflation}.
After that, we highlight the relation to the models based on Einstein frame supergravity, taking the $\alpha$-attractor models for specific examples.

The difference in the lower limit on the tensor-to-scalar ratio between two models shown in the previous subsections
originates from the difference in the coefficient $a_2$ of the leading pole in the kinetic term in the Einstein frame [see Eq.~\eqref{observables_p2}].
To see the appearance of such a difference, let us take a closer look at the structure of the kinetic term of the inflaton field.
From Eqs.~\eqref{conformal_factor} and~\eqref{condition_canonical_varphi}, 
we can rewrite the kinetic term of the inflaton field in the Einstein frame in terms of the conformal factor $\Omega^2$,
\begin{equation}
\frac{\mathcal{L}_{{\rm kin},E}}{\sqrt{-g_E}} = -\frac{1}{2}\left(\frac{1}{\Omega^2}+\frac{6(\Omega')^2}{\Omega^2}\right)(\partial\varphi)^2.
\end{equation}
Then, if we adopt the following relation between $\rho$ and $\varphi$,
\begin{equation}
\rho \propto \Omega^{-\frac{2}{q}}, \label{rho_Omega_relation}
\end{equation}
where $q$ is some real constant, 
the kinetic term reduces to
\begin{equation}
\frac{\mathcal{L}_{{\rm kin},E}}{\sqrt{-g_E}} = -\frac{1}{2}\left(\frac{q^2}{4(\Omega')^2}+\frac{3q^2}{2}\right)\frac{(\partial\rho)^2}{\rho^2}. \label{pole_q}
\end{equation}
If $\Omega'$ remains constant, we can regard the coefficient of $(\partial\rho)^2/\rho^2$ in the above equation
as a constant $a_2$. Hence, from Eq.~\eqref{observables_p2} we estimate the prediction for the tensor-to-scalar ratio as
\begin{equation}
r = \frac{8a_2}{N^2} = \frac{8}{N^2}\left(\frac{q^2}{4(\Omega')^2}+\frac{3q^2}{2}\right).
\end{equation}
It is straightforward to check that the conformal factor $\Omega^2$ satisfying the condition $\Omega'=\mathrm{const}.$ should be
a quadratic function of $\varphi$, and that both the FKLMP model and the $\tilde{\xi}$-model
in Sec.~\ref{sec:model_with_quadratic_frame_function} satisfy this condition.

If $(\Omega')^2>0$, the lower limit on $a_2$ is fixed by the second term $3q^2/2$,
which does not depend on the parameters such as $\xi$ and $\tilde{\xi}$, but on the value of $q$.
For instance, we have $q=1$ and $a_2 > 3/2$ in the FKLMP model, while $q=2$ and $a_2 > 6$ in the $\tilde{\xi}$-model.
This fact results in the different lower limit on the tensor-to-scalar ratio $r \gtrsim 48/N^2$ in the $\tilde{\xi}$-model, which is four times larger than that in the FKLMP model.

We note that the validity of the choice in Eq.~\eqref{rho_Omega_relation} depends on the form of the scalar potential in the Jordan frame.
Namely, we must guarantee that the function $f(\phi)$ is chosen such that the scalar potential in the Einstein frame takes the form~\eqref{condition_potential}
with $\rho$ given by Eq.~\eqref{rho_Omega_relation}.
In general, we can construct the desired form of $V_E(\rho)$ for $\rho = \Omega^{-2/q}$ with an arbitrary value of $q$ by choosing the scalar potential in the Jordan frame such that
\begin{equation}
V_J(\varphi) = \Omega(\varphi)^4 V_E(\rho) \quad \text{with} \quad \rho = \Omega(\varphi)^{-\frac{2}{q}}. \label{V_J_depends_on_q}
\end{equation}
The above equation explicitly shows that the choice of $V_J(\varphi)$ [or the choice of the function $f(\phi)$] depends on the value of $q$.
In other words, it is possible to construct a set of models that give rise to the same form of $V_E(\rho)$ but have different values of $q$
(or different lower limits on $r$) by adjusting the scalar potential appropriately in the Jordan frame.

The above discussion does not imply that the $\tilde{\xi}$-model constructed in the previous subsection is equivalent to the FKLMP model
except for the value of $q$. This is because the kinetic term and the scalar potential of these models are not exactly the same in the Einstein frame
although they are the same at leading order in $\rho$. 
More specifically, we have
\begin{equation}
K_E(\rho) = \left[\frac{1}{4\xi(1-\rho)}+\frac{3}{2}\right]\frac{1}{\rho^2}, \quad V_E(\rho) \propto (1-\rho)^2,
\end{equation}
for the FKLMP model with $\rho = \Omega^{-2}$, and
\begin{equation}
K_E(\rho) = \left[\frac{3}{\tilde{\xi}}+6\right]\frac{1}{\rho^2}, \quad V_E(\rho) \propto (1-\rho)^{2m},
\end{equation}
for the $\tilde{\xi}$-model with $\rho = \Omega^{-1}$.
The difference in the subleading terms leads to the different predictions for $n_s$ and $r$ as shown in Fig.~\ref{fig1}.
However, by adjusting the scalar potential in the Jordan frame
we can also construct a model that would lead to completely the same predictions except for the lower limit on $r$.
For instance, it is possible to construct a variant of the $\tilde{\xi}$-model
that would exhibit $q=1$ rather than $q=2$ by adjusting $f(\phi)$ according to Eq.~\eqref{V_J_depends_on_q}.
In such a case, $f(\phi)$ must be different from the monomial function $f = \lambda\phi^m$ specified in the previous subsection.

Next, let us consider the relation between the models constructed in the Jordan frame and those constructed in the Einstein frame,
particularly the $\alpha$-attractor models.

We can show that the K\"{a}hler potential of the $\tilde{\xi}$-model [Eq.~\eqref{Kahler_quad_model}] reduces to the following form (up to constant terms),
\begin{equation}
\mathcal{K} = -3\alpha \log(T+\bar{T}), \label{Kahler_alpha_attractor}
\end{equation}
if we identify
\begin{align}
T &= 1+\sqrt{\frac{2\tilde{\xi}}{3}}\phi, \\
\alpha &= 4\left(1+\frac{1}{2\tilde{\xi}}\right). \label{relation_alpha_tilde_xi}
\end{align}
Equation~\eqref{Kahler_alpha_attractor} corresponds to the K\"{a}hler potential for the $\alpha$-attractor models
with half-plane valuables $T$ and $\bar{T}$~\cite{Cecotti:2014ipa,Kallosh:2015zsa}.
In such models, the parameter $\alpha$ is interpreted as the scalar curvature of the K\"{a}hler manifold,
\begin{equation}
\mathcal{R}_K = - g_{T\bar{T}}^{-1}\partial_T\partial_{\bar{T}}\log g_{T\bar{T}} = -\frac{2}{3\alpha}. \label{Kahler_curvature}
\end{equation}
In the $\alpha$-attractor models defined in the Einstein frame, we can deduce the condition $\alpha>0$ from the requirement that
the kinetic term in the Einstein frame must have the proper sign.
Since the prediction for the tensor-to-scalar ratio is given by $r=12\alpha/N^2$~\cite{Kallosh:2013yoa,Carrasco:2015pla},
this fact implies that there is no lower limit on $r$ in the $\alpha$-attractor models.
On the other hand, in the $\tilde{\xi}$-model defined in the Jordan frame, 
the lower limit on $\alpha$ in Eq.~\eqref{relation_alpha_tilde_xi} becomes $\alpha>4$
as long as we require that $\tilde{\xi}$ should be real and positive.

It may appear surprising that two models (the $\tilde{\xi}$-model and the $\alpha$-attractor model)
that have exactly the same structure in the K\"{a}hler potential 
lead to different lower limits on the parameter $\alpha$
(or on the tensor-to-scalar ratio). This difference arises from a peculiar feature of the Jordan frame.
If we formally allow negative values of $\tilde{\xi}$ in Eq.~\eqref{relation_alpha_tilde_xi},
there would be no lower limit on $\alpha$ except for the condition $\alpha>0$ due to the requirement of the positive kinetic energy in the Einstein frame.
However, the conformal factor in Eq.~\eqref{Phi_tilde_xi}, $\Omega^2 = (1+(\tilde{\xi}/3)^{1/2}\varphi)^2$, becomes complex for $\tilde{\xi}<0$, 
which leads to the ill-defined relation between the Jordan frame and the Einstein frame.
Note that it is also possible to check this fact by using Eq.~\eqref{pole_q}.
Substituting $\Omega^2 = (1+(\tilde{\xi}/3)^{1/2}\varphi)^2$ and $q=2$ into Eq.~\eqref{pole_q}, we have
\begin{equation}
\frac{\mathcal{L}_{{\rm kin},E}}{\sqrt{-g_E}} = -\frac{1}{2}\left(\frac{3}{\tilde{\xi}}+6\right)\frac{(\partial\rho)^2}{\rho^2}. \nonumber
\end{equation}
Then, we see that there would be no lower limit on $a_2$ if we allow $\tilde{\xi}<0$.
This corresponds to the case with $(\Omega')^2<0$, which is incompatible with our assumption.

Unlike the $\tilde{\xi}$-model, the K\"{a}hler potential of the FKLMP model is not equivalent to that of the $\alpha$-attractor models,
but the peculiarity of the Jordan frame mentioned above is also relevant to the FKLMP model.
In this case, the conformal factor is given by $\Omega^2 = 1+\xi\varphi^2$.
Substituting it and $q=1$ into Eq.~\eqref{pole_q}, we have $a_2 \to (4\xi)^{-1}+3/2$ in the large field limit.
Then, if we formally allow negative values of $\xi$, there would be no lower limit on $a_2$.
Such a regime is forbidden since $\Omega^2$ vanishes for a finite value of $\varphi$, 
and we cannot define the appropriate conformal transformation.

To recap the above discussions, 
the $\tilde{\xi}$-model constructed in the previous subsection is inequivalent to the FKLMP model
in the sense that behaviors in the Einstein frame are not exactly the same, while
in general it is possible to construct plural models in the Jordan frame that lead to
the same dynamics in the Einstein frame by adjusting the scalar potential in the Jordan frame appropriately.
For such models, there exists a finite lower limit on the tensor-to-scalar ratio due to the restriction that
the conformal factor must be real and positive.
Furthermore, the value of the lower limit is determined by the choice of
the relation between $\rho$ and $\varphi$ [such as Eq.~\eqref{rho_Omega_relation}],
which originates from the choice of the scalar potential in the Jordan frame.

\section{Summary and discussions}
\setcounter{equation}{0}

In this paper, we have discussed inflation models in Jordan frame
supergravity. First of all, we have explicitly shown that the model
proposed by Ferrara {\it et al.} is uniquely characterized by the
following two requirements: (1) The kinetic terms of scalar fields
should be canonical in the Jordan frame. (2) The frame function consists
only of a holomorphic term (and its anti-holomorphic counterpart) for
symmetry breaking terms. Then, we tried to relax the latter condition
with keeping the former condition. In order to realize concrete examples
of inflation models with asymptotically flat potential in this context,
we introduced pole inflation technique and  
a stabilizer field. By
imposing the canonical kinetic terms and pole structure, we have derived
the condition for the frame function, 
whose solution yields a suitable example for 
inflation in Jordan frame supergravity. It should be
noticed that pole structure does not necessarily guarantee potential
suitable for inflation because it just guarantees asymptotic flatness of
the potential and the potential can be runaway type and/or have a bump,
which prevents the inflation from ending. We have confirmed that our
model can certainly realize inflation 
with the help of the stabilizer field. It is quite interesting to embed the Higgs field
and/or some fields beyond the Standard Model of particle physics to our
models, which is left for future work.

The model explored in Sec.~\ref{sec:model_with_quadratic_frame_function}
predicts that there is a lower limit on the tensor-to-scalar ratio, $r\gtrsim 0.013$. 
We emphasize that the precise value of this limit is model-dependent,
and closely related to the choice of the scalar potential in the Jordan frame, 
as shown in Sec.~\ref{sec:relation_to_kinetic_formulation}.
It is notable that, if the predicted value of $r$ is sufficiently large,
it is potentially accessible in future CMB $B$-mode experiments.  
Currently, the Planck collaboration has obtained a bound $r < 0.10$ (95\% CL, Planck TT+lowP) at the scale
$k=0.002\,\mathrm{Mpc}^{-1}$~\cite{Ade:2015lrj}, and it is expected that
the sensitivity on $r$ will be considerably improved in future ground-,
balloon-, and space-based experiments~\cite{Errard:2015cxa}.  Such
future experimental studies will test the model discussed in this paper.

We have explicitly shown that 
the K\"{a}hler potential of
the $\tilde{\xi}$-model in Sec.~\ref{sec:model_with_quadratic_frame_function}
is equivalent to 
that of the $\alpha$-attractor models, but there exists a non-trivial lower limit on the $\alpha$ parameter.
This fact originates from the peculiar feature of the Jordan frame
that the conformal factor must be real and positive such that
we can define the appropriate conformal transformation between the Jordan frame
and the Einstein frame. The choice of the frame function 
and the scalar potential in the Jordan frame
can constrain the parameter space of the equivalent model in the Einstein frame,
which leads to different predictions for the lower limit on the tensor-to-scalar ratio.

We emphasize that the model constructed in
Sec.~\ref{sec:model_with_quadratic_frame_function} is just one specific
example, in which three guiding principles enumerated in
Sec.~\ref{sec:beyond_FKLMP_formulation} are satisfied.  This fact does
not exclude the possibility of obtaining other models that lead to pole
inflation with the scalar fields having canonical kinetic terms in the
Jordan frame. It would be interesting to extend our analysis to explore
such inflationary models in a generic manner.\footnote{A model defined in Eq.~(5.6) of Ref.~\cite{Cecotti:2014ipa}
can be regarded as another example that contains non-holomorphic terms in the frame function
and leads to the same inflationary cosmology as the $\alpha$-attractor models. 
Such a model is inequivalent to that discussed in Sec.~\ref{sec:model_with_quadratic_frame_function} in a sense that
the K\"{a}hler curvature is different from that shown in Eq.~\eqref{Kahler_curvature}.}

In addition to the inflation model with the quadratic frame function
obtained in Sec.~\ref{sec:model_with_quadratic_frame_function}, we also
found that naive non-holomorphic extensions of the CSS models with
single power law functions lead to the scalar potential exhibiting
runaway behavior, which is not suitable for describing inflation
(see Sec.~\ref{sec:generalize_CSS} and
Appendix~\ref{sec:non_holomorphic_Delta_CSS}). However, it can be used to
construct a model that accounts for the accelerated expansion of the
present Universe (see Ref.~\cite{Ratra:1987rm} for example), though
freezing model is now less favored observationally. Here, the challenge
is to construct the model compatible with the low energy supersymmetry
breaking, which deserves further investigation.

\section*{Acknowledgments}
We would like to thank Renata Kallosh and Andrei Linde for discussions and important information.
M.Y. also thanks to Takeshi Chiba for useful comments.
This work was supported in part by JSPS Grant-in-Aid for Scientific
Research Nos.~25287054 (M.Y.)  and 26610062 (M.Y.), MEXT Grant-in-Aid
for Scientific Research on Innovative Areas ``Cosmic Acceleration''
No. 15H05888 (M.Y.), JSPS Postdoctoral Fellowships for Research Abroad (D.Y.).

\appendix

\section{Non-holomorphic $\Delta$-deformed CSS models}
\setcounter{equation}{0}
\label{sec:non_holomorphic_Delta_CSS}

In this appendix, we consider a non-holomorphic deformation of the CSS models discussed in Sec.~\ref{sec:generalize_CSS}.
Here we adopt the following K\"ahler potential and frame function
\begin{align}
\mathcal{K} &= -3\log\left(-\frac{1}{3}\frac{\Phi}{Y}\right), \\
\Phi &= -3 + \delta_{\alpha\bar{\beta}}z^{\alpha}\bar{z}^{\bar{\beta}} + \Delta_1 \mathcal{G}(z,\bar{z}), \\
Y&= 1 + \Delta_2 \mathcal{Y}(z,\bar{z}),
\end{align}
where $\mathcal{G}$ and $\mathcal{Y}$ are non-holomorphic functions,
and $\Delta_1$ and $\Delta_2$ are some real constant parameters.
From Eqs.~\eqref{L_kin_J} and~\eqref{Kahler _metric_Phi_Y}, the kinetic terms in the Jordan frame read
\begin{align}
\frac{\mathcal{L}_{{\rm kin},J}}{\sqrt{-g_J}} = - \left[\delta_{\alpha\bar{\beta}}+\Delta_1\mathcal{G}_{\alpha\bar{\beta}}
+\Delta_2\frac{\Phi}{Y^2}\left(\Delta_2 \mathcal{Y}_{\alpha}\mathcal{Y}_{\bar{\beta}}-Y\mathcal{Y}_{\alpha\bar{\beta}}\right)\right]g_J^{\mu\nu}\hat{\partial}_{\mu} z^{\alpha}\hat{\partial}_{\nu}\bar{z}^{\bar{\beta}},
\end{align}
where the subscripts $\alpha$ and $\bar{\beta}$ represent a derivative with respect to $z^{\alpha}$ and $\bar{z}^{\bar{\beta}}$, respectively.
If we take $\Delta_1 = \Delta_2 = 0$, the kinetic terms in the Jordan frame are canonically normalized,
and the CSS models are recovered.

Hereafter, we focus on the models with two real scalar fields $\phi$ and $S$ with the superpotential given by Eq.~\eqref{W_function}.
If we further assume that $S$ is stabilized at $S=0$ and that derivatives of $\mathcal{G}$ and $\mathcal{Y}$ with respect to $S$ and $\bar{S}$
vanish at that point, we can still use Eqs.~\eqref{V_E_stabilized} and~\eqref{gSS_stabilized}, and the scalar potential in the Einstein frame is given by
\begin{equation}
V_E = \frac{9Y^3}{\Phi^2}\left|f\right|^2 = \frac{(1+\Delta_2\mathcal{Y})^3}{(1-\frac{1}{3}|\phi|^2-\frac{\Delta_1}{3}\mathcal{G})^2}\left|f\right|^2.
\label{V_E_non_holomorphic_Delta_CSS}
\end{equation}

Let us assume that $f$, $\mathcal{G}$, and $\mathcal{Y}$ are given by single power law functions of $\phi$ at $S=0$,
\begin{equation}
f(\phi) = \lambda \phi^l,\quad \mathcal{G}(\phi,\bar{\phi}) = |\phi|^{2m}, \quad \text{and} \quad \mathcal{Y}(\phi,\bar{\phi}) = |\phi|^{2n},
\end{equation}
where $l$, $m$, and $n$ are positive integers, and $\lambda$ is a complex parameter.
Then, the scalar potential becomes a function of the radial direction $\varphi\equiv|\phi|/\sqrt{2}$,\footnote{The angular direction of $\phi$
can be stabilized by introducing an appropriate holomorphic function in $\Phi$~\cite{Nakayama:2010ga}.}
\begin{equation}
V_E = \frac{|\lambda|^2}{2^l}\frac{\left(1+\frac{\Delta_2}{2^n}\varphi^{2n}\right)^3}{(1-\frac{1}{6}\varphi^2-\frac{1}{3}\frac{\Delta_1}{2^m}\varphi^{2m})^2}\varphi^{2l}.
\label{V_E_power_law}
\end{equation}
In particular, it approaches to a constant value in the large field limit if $l=2m-3n$, which appears to be suitable for inflation.
However, in such cases the potential exhibits runaway behavior, as shown below.

In order to see the behavior of the scalar potential in the large field limit, we differentiate Eq.~\eqref{V_E_power_law}
with respect to $\varphi$ and impose the condition $l=2m-3n$,\footnote{In general, $\varphi$ has a non-canonical kinetic term 
in the Einstein frame, and the scalar field dynamics should be analyzed in terms of the canonically normalized field rather than $\varphi$.
However, the normalization of $\varphi$ does not affect the conclusion that the potential exhibits runaway behavior.}
\begin{align}
\frac{\partial V_E}{\partial\varphi} &= \left[2m-3n + \frac{1}{6}(2-2m+3n)\varphi^2+\frac{n\Delta_1}{2^m}\varphi^{2m}+\left(2m+\frac{1-m}{3}\varphi^2\right)\frac{\Delta_2}{2^n}\varphi^{2n}\right] \nonumber\\
&\quad \times \frac{2V_E}{\varphi\left(1+\frac{\Delta_2}{2^n}\varphi^{2n}\right)\left(1-\frac{1}{6}\varphi^2-\frac{1}{3}\frac{\Delta_1}{2^m}\varphi^{2m}\right)}.
\label{d_V_E_d_varphi}
\end{align}
We also note that $\Delta_1<0$ and $\Delta_2 > 0$ are required, since otherwise
we have $\Phi=0$ or $Y=0$ for a finite value of $\varphi$,
which leads to a singularity in the scalar potential or in the kinetic term, and prevent
$\varphi$ from rolling down to a small field value $\varphi \ll 1$ after inflation.
With these preliminaries, we can consider two different cases according to the value of $m$.
If $m>1$, the term with the highest power in the bracket in the first line of Eq.~\eqref{d_V_E_d_varphi}
is $n \Delta_1 |\phi|^{2m}$ or $(1-m) \Delta_2 |\phi|^{2n+2}/3$,
and both of them are negative, since we assume that $\Delta_1<0$ and $\Delta_2 > 0$.
Furthermore, the denominator of the second line of Eq.~\eqref{d_V_E_d_varphi} approaches to
$- \sqrt{2} \Delta_1 \Delta_2 |\phi|^{2(n+m)+1}/3$ in the large field limit, and it is positive.
Therefore, $\partial V_E/\partial\varphi$ becomes negative in the large field limit, which implies that 
the potential exhibits runaway behavior for $m>1$.
On the other hand, if $m=1$, which is only possible for $l=2$ and $n=0$
since we assume that $l=2m-3n$ and that $l$, $m$, $n$ are positive integers,
Eq.~\eqref{d_V_E_d_varphi} reduces to
\begin{equation}
\frac{\partial V_E}{\partial\varphi} = \frac{4V_E}{\varphi\left(1-\frac{1}{6}(1+\Delta_1)\varphi^2\right)}. \nonumber
\end{equation}
This becomes positive for $\Delta_1<-1$, but in such a case the kinetic term in the Einstein frame has a wrong sign,
$g_{\phi\bar{\phi}}=(1+\Delta_1)/(1-\frac{1}{6}(1+\Delta_1)\varphi^2)^2 < 0$.
From these facts, we conclude that the potential~\eqref{V_E_power_law} always exhibits runaway behavior in the large field limit
when $l=2m-3n$.

\section{Stabilization of non-inflaton fields}
\setcounter{equation}{0}
\label{sec:stabilization}

In this appendix, we show that the K\"{a}hler
potential, frame function, and superpotential
given by Eqs.~\eqref{Kahler_quad_model}-\eqref{non_horomorphic_J_quad_model}
are compatible with the assumption that ${\rm Im}(\phi)=0$ and $S=0$ during
inflation if we introduce $S$-dependence in the frame function and the
K\"{a}hler potential appropriately.

For simplicity, we assume that the frame function and the K\"ahler potential depend only on $|S|^2$, 
not on $S^2$ or $\bar{S}^2$, 
and that the function $Y$ is independent of $|S|^2$. 
Therefore, we use the same expression for $Y$ as in Eq.~\eqref{Y_quad_model}, 
which is rewritten in terms of $\phi$ and $\bar{\phi}$ as 
\begin{align} \label{Y_xi}
 \log Y = -2\left( 1+ {1\over \tilde{\xi}}  \right) \log \left( 1+ \sqrt{{\tilde{\xi} \over 6}} (\phi +\bar{\phi}) \right)\,.
\end{align}
On the other hand, we consider the following frame function,
\begin{align} \label{Phi_S}
 \Phi= \mathrm{e}^{F(|S|^4)} \left[ -3 \left(1+\sqrt{\tilde{\xi} \over 6} (\phi + \bar{\phi}) \right)^2 + |S|^2 \right]\,,
\end{align}
where $F(|S|^4)$ is introduced in order to guarantee the stabilization of $S$,
and it is a function of $|S|^4$ satisfying $F \rightarrow 0$ for $S \rightarrow 0$.
This frame function and the function \eqref{Y_xi} satisfy the conditions 
for which the kinetic terms of $\phi$ and $S$ become canonical,
\begin{align} \label{canonical_cond}
 \left( \log Y \right)_{\alpha \bar{\beta}} = { \Phi_{\alpha \bar{\beta}} -\delta_{\alpha\bar{\beta}} \over \Phi}\,, 
\end{align}
when $S=0$. 
From Eqs.~\eqref{Y_xi} and \eqref{Phi_S}, the K\"{a}hler potential can be written as 
\begin{align}\label{Kahler_S}
 \mathcal{K} = -3 \log \left[ \left(1+\sqrt{\tilde{\xi} \over 6} (\phi + \bar{\phi}) \right)^2 - {1\over3}|S|^2 \right] 
 -6\left( 1+ {1\over \tilde{\xi}}  \right) \log \left( 1+ \sqrt{{\tilde{\xi} \over 6}} (\phi +\bar{\phi}) \right) 
 -3F(|S|^4)\,.
\end{align}

Using the frame function \eqref{Phi_S} and the K\"ahler potential \eqref{Kahler_S}, 
we calculate the effective mass for ${\rm Im}(\phi)$ and $S$: 
\begin{align}
 &m^2_{{\rm Im}(\phi)} \equiv {1\over2} \mathcal{K}_{\phi \bar{\phi}}^{-1} 
 \left. {\partial^2 V_E \over \partial \left({\rm Im}(\phi) \right)^2 } \right|_{{\rm Im}(\phi)=S=0}\,,  \label{mass_Imphi}\\ 
 &m^2_S \equiv \mathcal{K}_{S \bar{S}}^{-1} 
 \left. {\partial^2 V_E \over \partial s^2 } \right|_{{\rm Im}(\phi)=S=0}\,,\label{mass_S}
 \end{align}
where $s = \sqrt{2}|S|$. 
Because the frame function and the K\"ahler potential depend only on $|S|^2$, 
the mass for ${\rm Re}(S)$ and ${\rm Im}(S)$ are equivalent, 
which can be easily confirmed from the detailed calculation that we will show in the following. 

In order to compute the masses~\eqref{mass_Imphi} and~\eqref{mass_S}, 
we need to compute $\mathcal{K}_\alpha$ and $\mathcal{K}_{\alpha \bar{\beta}}$. 
The first derivatives of the K\"ahler potential become 
\begin{align}
 &\partial_\phi \mathcal{K}= \partial_{\bar{\phi}} \mathcal{K}
 = - \sqrt{{6\over \tilde{\xi}}} {1+2\tilde{\xi} \over 1+ \sqrt{{\tilde{\xi} \over 6}} (\phi +\bar{\phi})} + \mathcal{O}(S^2)\,, \\
 &\partial_S \mathcal{K}= \overline{\partial_{\bar{S}} \mathcal{K} }
 =\bar{S} \left( 1+ \sqrt{{\tilde{\xi} \over 6}} (\phi +\bar{\phi}) \right)^{-2} + \mathcal{O}(S^3) \,. 
\end{align}
The second derivatives of the K\"ahler potential are given as 
\begin{align}
 &\partial_\phi \partial_{\bar{\phi}} \mathcal{K}
  = (1+2\tilde{\xi})  \left( 1+ \sqrt{{\tilde{\xi} \over 6}} (\phi +\bar{\phi}) \right)^{-2} + \tilde{\xi}  \left( 1+ \sqrt{{\tilde{\xi} \over 6}} (\phi +\bar{\phi}) \right)^{-4} |S|^2 + \mathcal{O}(S^3)\,, \label{K_phi_barphi}\\
 &\partial_S \partial_{\bar{\phi}} \mathcal{K}= \overline{\partial_\phi \partial_{\bar{S}} \mathcal{K}}
 = -2 \sqrt{{\tilde{\xi} \over 6}}  \left( 1+ \sqrt{{\tilde{\xi} \over 6}} (\phi +\bar{\phi}) \right)^{-3} \bar{S} + \mathcal{O}(S^3) \,, \\
 &\partial_S \partial_{\bar{S}} \mathcal{K}
  =   \left( 1+ \sqrt{{\tilde{\xi} \over 6}} (\phi +\bar{\phi}) \right)^{-2} 
  + {2\over3}\left( 1+ \sqrt{{\tilde{\xi} \over 6}} (\phi +\bar{\phi}) \right)^{-4} |S|^2
  - 3F_{S\bar{S}} + \mathcal{O}(S^3) \,, 
\end{align}
where $F_{S\bar{S}} = \partial_S \partial_{\bar{S}}F$. 
Therefore, $g^{\alpha \bar{\beta}}$ become
\begin{align}
&g^{\phi \bar{\phi}} = {1 \over 1+2\tilde{\xi}} \left( 1+ \sqrt{{\tilde{\xi} \over 6}} (\phi +\bar{\phi}) \right)^{2} 
+ \mathcal{O}(S^2)\,, \\
&g^{\phi \bar{S}} = \overline{g^{S \bar{\phi}}} = { 2 \sqrt{{\tilde{\xi} \over 6}} \over 1+2\tilde{\xi}} \left( 1+ \sqrt{{\tilde{\xi} \over 6}} (\phi +\bar{\phi}) \right) \bar{S} + \mathcal{O}(S^3)\,, \\
&g^{S \bar{S}} =  \left( 1+ \sqrt{{\tilde{\xi} \over 6}} (\phi +\bar{\phi}) \right)^2
+ \left[ {3F_{S\bar{S}} \over |S|^2}  \left( 1+ \sqrt{{\tilde{\xi} \over 6}} (\phi +\bar{\phi}) \right)^4  - {2(1+\tilde{\xi}) \over 3(1+2\tilde{\xi})} \right] |S|^2 + \mathcal{O}(S^3)\, .
\end{align}
$V_E$ is given as 
\begin{align}
 V_E &= {\mathrm e}^{\mathcal{K}} 
 \left( g^{\alpha \bar{\beta}} \nabla_\alpha W \nabla_{\bar{\beta}} \bar{W} -3|W|^2 \right) \nonumber\\
 &= |\lambda|^2 |\phi|^{2m}\left( 1+ \sqrt{{\tilde{\xi} \over 6}} (\phi +\bar{\phi}) \right)^{-2m}
 + |\lambda|^2 |S|^2 |\phi|^{2(m-1)}\left( 1+ \sqrt{{\tilde{\xi} \over 6}} (\phi +\bar{\phi}) \right)^{-2(m+1)} \nonumber\\
 & \qquad \times \left[ 2\left( {2+5\tilde{\xi} \over 3(1+2\tilde{\xi})} + {3\over \tilde{\xi}} +3 +{3F_{S\bar{S}}  \over2 |S|^2}   \left( 1+ \sqrt{{\tilde{\xi} \over 6}} (\phi +\bar{\phi}) \right)^4 \right) |\phi|^2 
 + {m^2 \over 1+2\tilde{\xi}} \left( 1-{\tilde{\xi} \over6} (\phi +\bar{\phi})^2 \right) \right]\,,
 \label{V_E_stabilization_full}
\end{align}
up to $\mathcal{O}(|S|^2)$. 
Setting $\phi=\bar{\phi} =\varphi/\sqrt{2}$ and $S=0$, we have
\begin{equation}
V_E = |\lambda|^2\left(\frac{\frac{\varphi}{\sqrt{2}}}{1+\sqrt{\frac{\tilde{\xi}}{3}}\varphi}\right)^{2m}.
\end{equation}
Therefore, the Hubble parameter during inflation in the large field limit $\varphi\gg\sqrt{3/\tilde{\xi}}$ reads
\begin{equation}
H_{\rm inf}^2 = \frac{V_E}{3} \simeq \left(\frac{3}{2\tilde{\xi}}\right)^m\frac{|\lambda|^2}{3}.
\end{equation}

From Eqs.~\eqref{mass_Imphi},~\eqref{K_phi_barphi}, and~\eqref{V_E_stabilization_full}, we obtain
\begin{equation}
m^2_{{\rm Im}(\phi)} = \frac{2m}{1+2\tilde{\xi}}\frac{|\lambda|^2}{2^m}\frac{\varphi^{2(m-1)}}{\left(1+\sqrt{\frac{\tilde{\xi}}{3}}\varphi\right)^{2(m-1)}} \\
 \simeq \frac{2(5\tilde{\xi}+3)}{1+2\tilde{\xi}}H_{\rm inf}^2\,,
\end{equation}
where we take the large field limit in the last equality. 
In this limit, $m^2_{{\rm Im}(\phi)} $ is larger than $H_{\rm inf}^2$ for $\tilde{\xi} > -5/8$.
Therefore, we confirm that the imaginary part of $\phi$ is stabilized at the origin during inflation.

Setting $F(|S|^4) = \zeta |S|^4/4$ and $\phi=\bar{\phi} =\varphi/\sqrt{2} $ in Eq.~\eqref{V_E_stabilization_full}, we have 
\begin{align}
 V_E &=  {|\lambda|^2 \over 2^m} \varphi^{2m}\left( 1+ \sqrt{{\tilde{\xi} \over 3}} \varphi  \right)^{-2m}
 + {|\lambda|^2 \over 2^{m-1}} |S|^2 \varphi^{2(m-1)}\left( 1+ \sqrt{{\tilde{\xi} \over 3}} \varphi \right)^{-2(m+1)} \nonumber\\
 & \qquad \times \left[ \left( {2+5\tilde{\xi} \over 3(1+2\tilde{\xi})} + {3\over \tilde{\xi}} +3 +{3\zeta \over2}   \left( 1+ \sqrt{{\tilde{\xi} \over 3}} \varphi \right)^4 \right) \varphi^2 
 + {m^2 \over 1+2\tilde{\xi}} \left( 1-{\tilde{\xi} \over3} \varphi^2 \right) \right] \,.
\end{align}
Then, $m_S^2$ becomes
\begin{align}
 m_S^2= {|\lambda|^2 \over 2^m} \varphi^{2(m-1)}\left( 1+ \sqrt{{\tilde{\xi} \over 3}} \varphi \right)^{-2m} \left[ \left(-{1\over3} +{3\zeta \over2}   \left( 1+ \sqrt{{\tilde{\xi} \over 3}} \varphi \right)^4 \right) \varphi^2 + {m^2 \over 1+2\tilde{\xi}} \right]\,,
\end{align}
where we used $m=5+3/\tilde{\xi}$. 
This clarifies that $m_S^2$ inevitably becomes negative for large $\varphi$ when $\zeta=0$ (or $F=0$). 
When $\zeta \neq 0$, the ratio between $m_S^2$ and $H_{\rm inf}^2$ becomes 
\begin{align}
 {m_S^2 \over H_{\rm inf}^2 } \sim \zeta \tilde{\xi}^2 \varphi^4 \,,
\end{align}
for large $\varphi$. 
Therefore, $S$ is sufficiently stabilized during inflation for positive $\zeta$. 



\begin{thebibliography}{99}

\bibitem{Bennett:2012zja} 
  C.~L.~Bennett {\it et al.}  [WMAP Collaboration],
  Astrophys.\ J.\ Suppl.\  {\bf 208}, 20 (2013)
  [arXiv:1212.5225 [astro-ph.CO]].

\bibitem{Adam:2015rua} 
  R.~Adam {\it et al.} [Planck Collaboration],
  Astron.\ Astrophys.\  {\bf 594}, A1 (2016)
  [arXiv:1502.01582 [astro-ph.CO]].
  
\bibitem{Hinshaw:2012aka} 
  G.~Hinshaw {\it et al.}  [WMAP Collaboration],
  Astrophys.\ J.\ Suppl.\  {\bf 208}, 19 (2013)
  [arXiv:1212.5226 [astro-ph.CO]].

\bibitem{Ade:2015lrj} 
  P.~A.~R.~Ade {\it et al.} [Planck Collaboration],
  Astron.\ Astrophys.\  {\bf 594}, A20 (2016)
  [arXiv:1502.02114 [astro-ph.CO]].

\bibitem{Starobinsky:1980te} 
  A.~A.~Starobinsky,
  Phys.\ Lett.\  {\bf 91B}, 99 (1980).

\bibitem{CervantesCota:1995tz} 
  J.~L.~Cervantes-Cota and H.~Dehnen,
  Nucl.\ Phys.\ B {\bf 442}, 391 (1995)
  [astro-ph/9505069].

\bibitem{Bezrukov:2007ep} 
  F.~L.~Bezrukov and M.~Shaposhnikov,
  Phys.\ Lett.\ B {\bf 659}, 703 (2008)
  [arXiv:0710.3755 [hep-th]].

\bibitem{Ferrara:2013rsa} 
  S.~Ferrara, R.~Kallosh, A.~Linde and M.~Porrati,
  Phys.\ Rev.\ D {\bf 88}, no. 8, 085038 (2013)
  [arXiv:1307.7696 [hep-th]].

\bibitem{Kallosh:2013yoa}
  R.~Kallosh, A.~Linde and D.~Roest,
  JHEP {\bf 1311} (2013) 198
  [arXiv:1311.0472 [hep-th]].

\bibitem{Carrasco:2015pla} 
  J.~J.~M.~Carrasco, R.~Kallosh and A.~Linde,
  JHEP {\bf 1510}, 147 (2015)
  [arXiv:1506.01708 [hep-th]].

\bibitem{Pallis:2013yda} 
  C.~Pallis,
  JCAP {\bf 1404}, 024 (2014)
  [arXiv:1312.3623 [hep-ph]].

\bibitem{Giudice:2014toa}
  G.~F.~Giudice and H.~M.~Lee,
  Phys.\ Lett.\ B {\bf 733} (2014) 58
  [arXiv:1402.2129 [hep-ph]].

\bibitem{Pallis:2014dma} 
  C.~Pallis,
  JCAP {\bf 1408}, 057 (2014)
  [arXiv:1403.5486 [hep-ph]].

\bibitem{Kallosh:2014rha} 
  R.~Kallosh,
  Phys.\ Rev.\ D {\bf 89}, no. 8, 087703 (2014)
  [arXiv:1402.3286 [hep-th]].

\bibitem{Galante:2014ifa} 
  M.~Galante, R.~Kallosh, A.~Linde and D.~Roest,
  Phys.\ Rev.\ Lett.\  {\bf 114}, no. 14, 141302 (2015)
  [arXiv:1412.3797 [hep-th]].

\bibitem{Broy:2015qna} 
  B.~J.~Broy, M.~Galante, D.~Roest and A.~Westphal,
  JHEP {\bf 1512}, 149 (2015)
  [arXiv:1507.02277 [hep-th]].

\bibitem{Terada:2016nqg} 
  T.~Terada,
  Phys.\ Lett.\ B {\bf 760}, 674 (2016)
  [arXiv:1602.07867 [hep-th]].

\bibitem{Nakayama:2016eqv} 
  K.~Nakayama, K.~Saikawa, T.~Terada and M.~Yamaguchi,
  JHEP {\bf 1605}, 067 (2016)
  [arXiv:1603.02557 [hep-th]].
  
\bibitem{Kobayashi:2017qhk} 
  T.~Kobayashi, O.~Seto and T.~H.~Tatsuishi,
  arXiv:1703.09960 [hep-th].
  
\bibitem{Maeda:1988ab} 
  K.~Maeda,
  Phys.\ Rev.\ D {\bf 39}, 3159 (1989).

\bibitem{ArmendarizPicon:2002qb} 
  C.~Armendariz-Picon,
  Phys.\ Rev.\ D {\bf 66}, 064008 (2002)
  [astro-ph/0205187].

\bibitem{Flanagan:2004bz} 
  E.~E.~Flanagan,
  Class.\ Quant.\ Grav.\  {\bf 21}, 3817 (2004)
  [gr-qc/0403063].

\bibitem{Catena:2006bd} 
  R.~Catena, M.~Pietroni and L.~Scarabello,
  Phys.\ Rev.\ D {\bf 76}, 084039 (2007)
  [astro-ph/0604492].

\bibitem{Deruelle:2010ht} 
  N.~Deruelle and M.~Sasaki,
  Springer Proc.\ Phys.\  {\bf 137}, 247 (2011)
  [arXiv:1007.3563 [gr-qc]].

\bibitem{Chiba:2013mha} 
  T.~Chiba and M.~Yamaguchi,
  JCAP {\bf 1310}, 040 (2013)
  [arXiv:1308.1142 [gr-qc]].

\bibitem{Kamenshchik:2014waa} 
  A.~Y.~Kamenshchik and C.~F.~Steinwachs,
  Phys.\ Rev.\ D {\bf 91}, no. 8, 084033 (2015)
  [arXiv:1408.5769 [gr-qc]].

\bibitem{Futamase:1987ua} 
  T.~Futamase and K.~Maeda,
  Phys.\ Rev.\ D {\bf 39}, 399 (1989).

\bibitem{Kallosh:2000ve} 
  R.~Kallosh, L.~Kofman, A.~D.~Linde and A.~Van Proeyen,
  Class.\ Quant.\ Grav.\  {\bf 17}, 4269 (2000)
  Erratum: [Class.\ Quant.\ Grav.\  {\bf 21}, 5017 (2004)]
  [hep-th/0006179].

\bibitem{Ferrara:2010yw} 
  S.~Ferrara, R.~Kallosh, A.~Linde, A.~Marrani and A.~Van Proeyen,
  Phys.\ Rev.\ D {\bf 82}, 045003 (2010)
  [arXiv:1004.0712 [hep-th]].

\bibitem{Ferrara:2010in} 
  S.~Ferrara, R.~Kallosh, A.~Linde, A.~Marrani and A.~Van Proeyen,
  Phys.\ Rev.\ D {\bf 83}, 025008 (2011)
  [arXiv:1008.2942 [hep-th]].

\bibitem{Kallosh:2013wya} 
  R.~Kallosh and A.~Linde,
  JCAP {\bf 1306}, 027 (2013)
  [arXiv:1306.3211 [hep-th]].

\bibitem{Freedman:2012zz} 
  D.~Z.~Freedman and A.~Van Proeyen,
  ``Supergravity,''
  Cambridge University Press, Cambridge U.K. (2012).  

\bibitem{Buchmuller:2012ex} 
  W.~Buchm\"uller, V.~Domcke and K.~Schmitz,
  JCAP {\bf 1304}, 019 (2013)
  [arXiv:1210.4105 [hep-ph]].
  
\bibitem{Das:2016kwz} 
  K.~Das, V.~Domcke and K.~Dutta,
  JCAP {\bf 1703}, no. 03, 036 (2017)
  [arXiv:1612.07075 [hep-ph]].
  
\bibitem{Nakayama:2010ga} 
  K.~Nakayama and F.~Takahashi,
  JCAP {\bf 1011}, 039 (2010)
  [arXiv:1009.3399 [hep-ph]].

\bibitem{Ford:1986sy} 
  L.~H.~Ford,
  Phys.\ Rev.\ D {\bf 35}, 2955 (1987).
  
\bibitem{Peebles:1998qn} 
  P.~J.~E.~Peebles and A.~Vilenkin,
  Phys.\ Rev.\ D {\bf 59}, 063505 (1999)
  [astro-ph/9810509].

\bibitem{Felder:1999pv} 
  G.~N.~Felder, L.~Kofman and A.~D.~Linde,
  Phys.\ Rev.\ D {\bf 60}, 103505 (1999)
  [hep-ph/9903350].
  
\bibitem{Kawasaki:2000yn} 
  M.~Kawasaki, M.~Yamaguchi and T.~Yanagida,
  Phys.\ Rev.\ Lett.\  {\bf 85}, 3572 (2000)
  [hep-ph/0004243].

\bibitem{Kallosh:2010xz} 
  R.~Kallosh, A.~Linde and T.~Rube,
  Phys.\ Rev.\ D {\bf 83}, 043507 (2011)
  [arXiv:1011.5945 [hep-th]].
  
\bibitem{Roest:2013aoa} 
  D.~Roest, M.~Scalisi and I.~Zavala,
  JCAP {\bf 1311}, 007 (2013)
  [arXiv:1307.4343 [hep-th]].

\bibitem{Cecotti:2014ipa} 
  S.~Cecotti and R.~Kallosh,
  JHEP {\bf 1405}, 114 (2014)
  [arXiv:1403.2932 [hep-th]].

\bibitem{Kallosh:2015zsa} 
  R.~Kallosh and A.~Linde,
  Comptes Rendus Physique {\bf 16}, 914 (2015)
  [arXiv:1503.06785 [hep-th]].
  
\bibitem{Errard:2015cxa} 
  J.~Errard, S.~M.~Feeney, H.~V.~Peiris and A.~H.~Jaffe,
  JCAP {\bf 1603}, no. 03, 052 (2016)
  [arXiv:1509.06770 [astro-ph.CO]].

\bibitem{Ratra:1987rm} 
  B.~Ratra and P.~J.~E.~Peebles,
  Phys.\ Rev.\ D {\bf 37}, 3406 (1988).
  
\end{thebibliography}
\end{document}